\documentclass[11pt,a4paper]{article}

\usepackage{amsmath,amsfonts,amssymb,amsthm,cite}

\evensidemargin=\oddsidemargin
\advance\topmargin by -\headheight
\advance\topmargin by -\headsep

\newtheorem{thm}{Theorem}[section]
\newtheorem{lem}[thm]{Lemma}

\theoremstyle{definition}               
\newtheorem{defn}[thm]{Definition}
\theoremstyle{remark}                   
\newtheorem{rem}[thm]{Remark}

\numberwithin{equation}{section}        

\renewcommand{\a}{\alpha}               
\newcommand{\Coo}{C^\infty}             
\newcommand{\Dslash}{{D\mkern-11.5mu/\,}} 
\DeclareMathOperator{\Diff}{Diff}
\newcommand{\ep}{\epsilon}

\newcommand{\Ga}{\Gamma}                
\newcommand{\ga}{\gamma}                
\renewcommand{\H}{\mathcal{H}}          
\newcommand{\HH}{\mathbb{H}}         

\DeclareMathOperator{\im}{Im}
\DeclareMathOperator{\Isom}{Isom}       
\renewcommand{\L}{\mathcal{L}}          
\newcommand{\Lie}{Lie}

\newcommand{\mop}{{\mathchoice{\mathbin{\star_{_\theta}}}
         {\mathbin{\star_{_\theta}}}       
         {{\star_\theta}}{{\star_\theta}}}}
\newcommand{\Mop}{{\mathchoice{\mathbin{\star_{_\Theta}}}
         {\mathbin{\star_{_\Theta}}}       
         {{\star_\Theta}}{{\star_\Theta}}}}
\newcommand{\N}{\mathbb{N}}             
\newcommand{\pa}{\partial}              
\newcommand{\Q}{\mathbb{Q}}             
\newcommand{\R}{\mathbb{R}}             

\newcommand{\sepword}[1]{\quad\mbox{#1}\quad} 
\newcommand{\Sf}{\mathbb{S}}            
\newcommand{\sihalf}{{\scriptstyle\frac{i}{2}}}
\DeclareMathOperator{\spec}{sp}         
\renewcommand{\SS}{\mathcal{S}}         
\DeclareMathOperator{\supp}{supp}       
\newcommand{\T}{\mathbb{T}}             

\renewcommand{\th}{\theta}              
\newcommand{\Th}{\Theta}
\newcommand{\thalf}{\tfrac{1}{2}}       
\DeclareMathOperator{\Tr}{Tr}           
\DeclareMathOperator{\tr}{tr}           
\DeclareMathOperator{\tsum}{{\textstyle\sum}} 
\newcommand{\tri}{\Delta}               

\newcommand{\vf}{\varphi}               
\newcommand{\x}{\times}                 
\newcommand{\Z}{\mathbb{Z}}             

\def\<#1,#2>{\langle#1,#2\rangle}       
\def\ee_#1{e_{{\scriptscriptstyle#1}}}  
\def\wick:#1:{\mathopen:#1\mathclose:}  

\begin{document}

\thispagestyle{empty}

\begin{center}

CENTRE DE PHYSIQUE TH\'EORIQUE$\,^1$\\
CNRS--Luminy, Case 907\\
13288 Marseille Cedex 9\\
FRANCE\\

\vspace{2cm}

{\LARGE \bf {Heat-Kernel Approach to UV/IR Mixing on
Isospectral Deformation Manifolds}} \\

\vspace{1.5cm}

{\large Victor Gayral$^2$} \\

\vspace{1.5cm}

{\large\textbf{Abstract}}
\end{center}
We work out the general features of perturbative field theory on
noncommutative manifolds defined by isospectral deformation. These (in general
curved) `quantum spaces', generalizing Moyal planes and noncommutative tori,
are constructed using Rieffel's theory of deformation quantization by actions
of~$\R^l$. Our framework, incorporating background field methods and tools of
QFT in curved spaces, allows to deal both with compact and non-compact spaces,
as well as with periodic and non-periodic deformations, essentially in the
same way. We compute the quantum effective action up to one loop for a scalar
theory, showing the different UV/IR mixing phenomena for different kinds of
isospectral deformations. The presence and behavior of the non-planar parts of
the Green functions is understood simply in terms of off-diagonal heat kernel
contributions. For periodic deformations, a Diophantine condition on the
noncommutivity parameters is found to play a role in the analytical nature of
the non-planar part of the one-loop reduced effective action. Existence of
fixed points for the action may give rise to a new kind of UV/IR mixing.

\vspace{1.5cm}


\vspace{1.5cm}

\noindent CPT-2004/P.135

\noindent $^1$ UMR 6207

-- Unit\'e Mixte de Recherche du CNRS et des
Universit\'es Aix-Marseille I,\\
Aix-Marseille II et de l'Universit\'e
du Sud Toulon-Var

-- Laboratoire affili\'e \`a la FRUMAM -- FR 2291\\
$^2$ Also at Universit\'e de Provence, gayral@cpt.univ-mrs.fr.\\

\section{Introduction}

Noncommutative geometry (NCG), specially in Connes' algebraic and operatorial
formulation~\cite{Book}, is an attempt to free oneself from the classical
differential structure framework in modeling and understanding space-time,
while keeping in algebraic form geometry's tools such as metric and spin
structures, vector bundles and connection theory. The NCG framework is well
adapted to deal with quantum field theory over `quantum' space-time
(NCQFT)~\cite{Atlas}. However, there is a lack of computable examples
crucially needed to progress in this direction. Here we present a large class
of models, the isospectral deformation manifolds, in which we show the
intrinsic nature of UV/IR mixing through the analysis of a scalar theory.

In~\cite{ConnesLa,ConnesDV} Connes, Landi and Dubois-Violette gave a method
to generate noncommutative spaces based on the noncommutative torus paradigm.
For any closed Riemannian spin (this last condition could be relaxed for our
purpose) manifold whith isometry group of rank $l\geq 2$, one can build a
family of noncommutative spaces, called isospectral deformations by the
authors. The terminology comes from the fact that the underlying spectral
triple, that is, the dual object $(\Coo(M_\Th),L^2(M,S),\Dslash)$ encoding
all the topological, differential, metric and spin structures of the original
manifold, and so defining the `quantum Riemannian' space~\cite{Connesgrav}, has the same
space of spinors and the same Dirac operator as the undeformed one
$(\Coo(M),L^2(M,S),\Dslash)$; only the algebra is modified.

More precisely, the noncommutative algebra
$\Coo(M_\Th)$ can be defined as a fixed point algebra under a
group action~\cite{ConnesDV}:
\begin{equation}
\label{fixedpoint}
\Coo(M_\Th):=\Big(\Coo(M)\widehat{\otimes}
\T^l_\Th\Big)^{\a\widehat{\otimes}\tau^{-1}},
\end{equation}
where $\T^l_\Th$ is a $l$-dimensional NC torus(-algebra) with deformation
matrix $\Th\in M_l(\R),\Th^t=-\Th;\;\a$ is the action of
$\T^l$ on $M$ given by an Abelian part of its isometry group, $\tau$ is the
standard action of~$\T^l$ on~$\T^l_\Th$ and~$\widehat{\otimes}$ is a suitable
tensor product completion. By the Myers-Steenrod Theorem\cite{MS},
which asserts that $\Isom(M,g)\subset SO(n)$ for any $n$-dimensional
compact Riemannian manifold $(M,g)$, one can see that the class
of such manifolds whose isometry group has rank greater or equal to two
is far from small.

V\'arilly~\cite{Larissa} and Sitarz~\cite{Sitarz} independently remarked that
this construction fits into Rieffel's theory of deformation quantization for
actions of $\R^l$~\cite{RieffelDefQ}. Given a Fr\'echet algebra $A$ with
seminorms $\{p_i\}_{i\in I}$ and a strongly continuous isometric (with
respect to each seminorms) action of $\R^l$, one can deform the product of the
subalgebra $A^\infty$, consisting of smooth elements of $A$ with respect to
the generators $X^k$, ${k\in\{1,\ldots,l\}}$ of the action~$\a$. The algebra
$A^\infty$ can be canonically endowed with a new set of seminorms
$\{\tilde{p}_{i,m}\}_{i\in I,m\in\N}$ given by $\tilde{p}_{i,m}(.):=
\sup_{j\leq i}\sum_{|\beta|\leq m} p_j(X^\beta.)$, $\beta\in\N^l$. Those seminorms have the
property of being compatible with the deformed product defined by the
$A^\infty$-valued oscillatory integral:
$$
a\Mop b:=(2\pi)^{-l}\int_{\R^{2l}}\,d^ly\,d^lz\;e^{-i<y,z>}\;
\a_{\thalf\Th y}(a)\a_{-z}(b),\;a,b\in A^\infty.
$$
Here $\Th$ is the (real, skewsymmetric) deformation $l\times l$
matrix, $<y,z>=\sum_{i=1}^ly^i\,z^i$, and if we denote by~$A^\infty_\Th$ the
algebra $(A^\infty,\Mop)$, the deformation process verifies
$(A^\infty_\Th)^\infty_{\Th'}=A^\infty_{\Th+\Th'}$, and hence is reversible. In
\cite{GIV}, we investigate the equivalent of~(\ref{fixedpoint}) in the non-periodic
case and extend the construction of isospectral non-periodic deformations (called also
$\th$-deformations to distinguish them from $q$-deformations) to non-compact
manifolds within Rieffel's framework, whose paradigms are now the Moyal
planes~\cite{Himalia}.

Although we will not use directly the fixed point
characterization~(\ref{fixedpoint}), we want to insist on its crucial
importance to understand the situation. Indeed, such a characterization means
that we are transferring the noncommutative structure of the NC torus or of
the Moyal plane inside the commutative algebra of smooth functions, in a way
compatible with the Riemannian structure.

The first studied examples of NCQFT were the NC tori and the Moyal planes, in
pioneer works like~\cite{CR, Filk, KW, MR, MRS, Atlas} (see also~\cite{DN}
and~\cite{Z} for reviews). In those flat space situations, the main novelty in
regard to renormalization aspects is that two kinds of Feynman diagrams
coexist, respectively called planar and non-planar. The first one yields
ordinary UV divergences, while the non-planar graphs, characterized by
vertices which depend on external momenta through a phase, are finite except
for some values of the incoming momenta. That happens in
particular for the zero mode in $\lambda \vf^{\Mop 4}$ theory on the NC torus
and in the limit $p^\mu\rightarrow 0$ for the same theory on the Moyal plane.
This is the famous UV/IR entanglement phenomenon, which gives rise to
difficulties for any renormalization scheme.

In this paper, we show that for any (in general non-flat) isospectral
deformation, UV/IR mixing in (Euclidean) NCQFT exists as in the (flat)
paradigmatic examples of the NC torus and the Moyal planes.

In the next section, isospectral deformations are constructed and their basic
NCG properties are reviewed. The third section is devoted to the study of the
$\lambda \vf^{\Mop 4}$ theory. One derives a field expansion from a (modified)
heat kernel asymptotics to compute the effective action up to one loop. This
construction gives a simple algebraic meaning to the presence and behavior of
planar and non-planar sectors in those theories. In sections 4 and 5, using
off-diagonal heat-kernel estimates, we prove the inherent generic character of the
divergent structures for all kinds of isospectral deformations.
Fixed points for the $\R^l$ action potentially yield a new kind of UV/IR mixing.

\section{Isospectral deformations}

As explained in the Introduction, isospectral deformations are curved
noncommutative spaces generalizing Moyal planes and noncommutative
torus. To construct those NC Riemannian spaces (spectral triples), we
use an approach developed in~\cite{GIV}. Advantages of this twisted
product approach \emph{\`a la} Rieffel are that it allows to treat on
the same footing compact and non-compact cases (unital and non-unital
algebras) as well as periodic and non-periodic deformations, and that
it is well adapted for Hilbertian analysis.

Let $(M,g)$ be a locally compact, complete, connected, oriented
Riemannian $n$-dimen\-sional manifold without boundary, and let $\a$ be
a smooth isometric action of $\R^l,\,2\leq l\leq n$
$$
\a:\R^l\longrightarrow \Isom(M,g)\subset\Diff(M),
$$
where $l$ is less or equal to the rank of the isometry group of
$(M,g)$. We can then define a deformed or twisted product. The
isometric action $\a$ yields a group of automorphisms on $C^\infty(M)$
that we will again denote by~$\a$: for all $z\in\R^l$
$$
\a_zf(p) := f(\a_{-z}(p)).
$$
For brevity we will often write $z.p\equiv\a_z(p)$ to designate
the action of a group element on a point of the manifold. Obviously, the group
action property reads
$$
z_1.(z_2.p)=(z_1+z_2).p\sepword{and} 0.p=p.
$$
The infinitesimal generators of this action
$$
X_j(.):=\left. \frac{\pa}{\pa z^j}\a_z(.)\right|_{z=0},\quad j=1,\cdots,l,
$$
are ordinary smooth vector fields, so they leave
$\Coo_c(M)$ invariant. Hence, given a real skewsymmetric $l\times l$ matrix
$\Th$, one defines the deformed product of any~$f,h\in C_c^\infty(M)$ as a
bilinear product on~$C_c^\infty(M)$ with values
in~$C^\infty(M)\cap L^\infty(M,\mu_g)$ by the
oscillatory integral
\begin{equation}
\label{mop}
f\Mop h:=(2\pi)^{-l}\int_{\R^{2l}}\,d^ly\,d^lz\;
e^{-i<y,z>}\;\a_{\thalf\Th y}(f)\a_{-z}(h),
\end{equation}
where $<y,z>:=\sum_{j=i}^ly^jz^j$ can be viewed as the pairing between $\R^l$ and its dual
group. In spite of appearances this formula is symmetric, even with a
degenerate $\Th$ matrix (see the discussion near the end of this
section), as one can rewrite the deformed product:
\begin{equation*}
f\Mop h:=(2\pi)^{-l}\int_{\R^{2l}}\,d^ly\,d^lz\;
e^{i<y,z>}\;\a_{-y}(f)\a_{\thalf\Th z}(h).
\end{equation*}
The non-locality of this product generates a non-preservation of
supports. In particular, the twisted product of two functions with
disjoint support turns out to be non-zero a priori. Whereas in the
periodic case ($\ker\a\simeq\Z^l$) the fixed point characterization
gives rise to a reasonable locally convex topology on the invariant
sub-algebra of the algebraic tensor product $\Big(\Coo(M)\otimes
\T^l_\Th\Big)^{\a\otimes\tau^{-1}}$ or $\Big(\Coo_c(M)\otimes
\T^l_\Th\Big)^{\a\otimes\tau^{-1}}$ depending whether $M$ is compact
or not, to obtain a smooth algebra structure in the non-periodic case
one has to complete $C_c^\infty(M)$ to a Fr\'echet algebra with
seminorms defined through the measure associated to the Riemannian
volume form, so that the action becomes strongly continuous and
isometric with respect to each seminorm. This feature is investigated
in \cite{GIV}. In the sequel, as we mainly work at the linear level,
$\Coo_c(M)$ will be deemed ``large enough''.

The associativity of the product~(\ref{mop}) can be easily checked.
The ordinary integral with Riemannian volume form $\mu_g$ is a trace
(a proof is provided in~\cite{GIV}):
\begin{equation}
\label{eq:trace}
\int_M\; \mu_g \;f\Mop h=\int_M\; \mu_g \;f\,h=\int_M\; \mu_g \;h\Mop
f;
\end{equation}
$\a$ is still an automorphism for the deformed product:
\begin{equation}
\label{eq:aut}
\a_z(f)\Mop \a_z(h)=\a_z(f\Mop h);
\end{equation}
the complex conjugation is an involution:
\begin{equation}
\label{eq:inv}
(f\Mop h)^*=h^*\Mop f^*;
\end{equation}
and the Leibniz rule is satisfied for the generators of the action
\begin{equation}
X^k(f\Mop h)=X^k(f)\Mop h+f\Mop X^k(h),\;k=1,\cdots,l.
\end{equation}
In fact, the Leibniz rule is satisfied for any order one differential
operator which commutes with the action $\a$, thus for the Dirac
operator when the manifold has a spin structure.

We have basically two distinct situations. When the group action is effective
($\ker\a=\{0\}$), i.e. for a non-periodic deformation, it is seen that the
good topological assumption on $\a$ in order to avoid serious difficulties is
properness. That is, we assume the map
$$
(z,p)\in\R^l\times M\mapsto (p,\a_z(p))\in M\times M
$$
to be proper. Recall that a map between topological spaces is proper
if the preimage of any compact set is compact as well. On the other
hand, for periodic deformations the action factors through a torus
action $\tilde{\a}:\R^l/\Z^l\to\Isom(M,g)$, and the factorized action
$\tilde{\a}$ is automatically proper.

When $M$ is compact, $\a$ must be periodic to be proper, while in the
non-compact case both situations appear. We point out that the (non-compact)
non-periodic case is the most difficult one. First, when the manifold is not
compact, the essential spectrum of the Laplacian is non-empty, so its negative
powers are no longer compact operators. Furthermore, for periodic deformations
(of compact manifolds or not) we have a spectral subspace decomposition,
indexed by the dual group of $\T^l$, which does simplify proofs and
computations. \\
We do not explicitly treat the mixed case
$\a:\R^d\times\T^{l-d}\to\Isom(M,g)$, but its general features will be clear
from what follows.

The hypothesis of geodesically completeness of $M$ guarantees
selfadjointness of the (closure of the) Laplace-Beltrami operator
$\tri$ restricted to (the dense subset $\Coo_c(M)$ of)~$L^2(M,\mu_g)$,
the separable Hilbert space of squared integrable functions with respect to the
measure space $(M,\mu_g)$. In our convention, $\tri=(d+\delta)^2$ is
positive, and reduced to 0-forms $\tri=\delta d=\ast_{_H}d\ast_{_H}d$
where $\ast_{_H}$ is the Hodge star. Completeness (plus boundedness
from below of the Ricci curvature) is needed to have conservation of
probability~\cite{Chavel,Davies}:
$$
\int_M\;\mu_g(p)\;K_t(p,p')=1,
$$
where $K_t:=K_{e^{-t\tri}}$ is the heat kernel of the manifold. Recall
that $K_t(p,p')$ for $t>0$ is a smooth strictly positive symmetric
function on $M\times M$. The restriction to manifolds without boundary
is required to have a simple (with vanishing of the odd terms
\cite{Gilkey}) on-diagonal expansion of the heat kernel
\begin{equation}
\label{eq:gilkey}
K_t(p,p)\simeq (4\pi
t)^{-n/2}\sum_{l\in\N}\;t^l\;a_{2l}(p),\;\;t\rightarrow 0,
\end{equation}
where $a_l(p)$ are the so called Seeley--De Witt coefficients.

It is proved in \cite{GIV} that for non-compact non-periodic
deformations (the statement being immediate in the periodic case)
$L_f\equiv L^\Th_f$ (resp. $R_f\equiv R^\Th_f$), the operator of left
(resp. right) twisted multiplication by $f$, defined by
$L_f\psi=f\Mop\psi$ (resp. $R_f\psi=\psi\Mop f$), for
$\psi\in\H:=L^2(M,\mu_g)$, is bounded for any $f\in C^\infty_c(M)$.
This will be also true for smooth functions decreasing fast enough at
infinity.

Denote by $V_z$ the induced action of $\R^l$ on $L^2(M,\mu_g)$ by
unitary operators
$$
V_z\psi(p) := \psi(-z.p);
$$
then one can alternatively define $L_f$ and $R_f$ by an operator
valued integral
\begin{equation}
\label{Lfint}
L_f=(2\pi)^{-l}\int_{\R^{2l}}\,d^ly\,d^lz\;
e^{-i<y,z>}\;V_{\thalf \Th y}\,M_f\,V_{-z},
\end{equation}
\begin{equation}
\label{Rfint}
R_f=(2\pi)^{-l}\int_{\R^{2l}}\,d^ly\,d^lz\;
e^{-i<y,z>}\;V_{-z}\,M_f\,V_{\thalf \Th y},
\end{equation}
where $M_f$ denotes the operator of pointwise
multiplication by $f$.\\
Such integrals do not define B\"ochner integrals in the vector space $\L(\H)$.
Indeed, the operatorial norm of the integrands in (\ref{Lfint}) and (\ref{Rfint})
are not integrable functions on $\R^{2l}$, since they
depend on $y$ and $z$ only through unitary operators. Actually, the latter
must be understood as $\L(\H)$-valued oscillatory integrals \cite{RieffelDefQ}.

Formulas (\ref{Lfint}) and (\ref{Rfint})
can be easily derived from (\ref{mop}) using
$$
V_zM_fV_{-z}=M_{\a_z(f)}
$$
and the translation $z\to z-\thalf\Th y$ which leaves invariant the
phase due to the skewsymmetry of the deformation
matrix. Note that they can be used to
define (left and right) `Moyal multiplications' of any bounded
operator on~$\H$, taking the place of~$M_f$ in the formulas. Within
this presentation, it is straightforward to check that $L$ and $R$ are
two commuting representations (in fact $R$ is an anti-representation):
$$
[L_f,R_h]=0,\;\;\forall f,h\in C^\infty_c(M).
$$
Thus formulas (\ref{Lfint}) and (\ref{Rfint}) provide an other way to
check the associativity of the twisted product, which is equivalent to the
commutativity of the left and right regular representations.

Using the trace property (\ref{eq:trace}), one can also prove that the
adjoint of the left (resp. right) twisted multiplication by $f$ equals
the left (resp. right) twisted multiplication by the complex conjugate
of $f$:
$$
(L_f)^*=L_{f^*},\,\,\,\,(R_f)^*=R_{f^*}.
$$
Again, this fact can be directly checked using formulas~(\ref{Lfint})
and~(\ref{Rfint}). For $L_f$ it reads
\begin{align*}
(L_f)^*=&(2\pi)^{-l}\int_{\R^{2l}}\,d^ly\,d^lz\;
e^{i<y,z>}\;V_{z}\,M_{f^*}\,V_{-\thalf\Th y}\\
=&(2\pi)^{-l}\int_{\R^{2l}}\,d^ly\,d^lz\;
e^{-i<y,z>}\;V_{\thalf \Th z}\,M_{f^*}\,V_{-y},
\end{align*}
where the changes of variable $z\to \thalf\Th z,y\to 2\Th^{-1}y$
and the relation $<\Th^{-1}y,\Th z>=-<y,z>$ have been used.

The primary example of such a space is the n-dimensional Moyal
plane~$\R^n_\Th$. In this case, the manifold is the flat Euclidean
space~$\R^n,l=n$, and $\R^n$ acts on itself by translation. Another interesting
non-compact space which carries a smooth action of~$\R^{n-1}$ by isometry is
the $n$-dimensional hyperbolic space $\HH^n$, that we can make into
noncommutative $\HH^n_\Th$ by the previous prescription.

For periodic actions, there is a lattice $L=\beta\Z^l, \beta\in
M_l(\Z)$ in the kernel of $\a$ which factors through a torus
$\T_\beta^l:=\R^l/\beta\Z^l$. This quotient is a compact space if and
only if the rank of $\beta$ equals $l$. In this case, we have a
spectral subspace (Peter-Weyl) decomposition (see \cite{ConnesLa,
RieffelDefQ, Larissa} for details): for any bounded operator $A$ which
is $\a$-norm smooth (the map $z\in\T_\beta^l\mapsto V_zAV_{-z}$
is smooth for the norm topology of $\L(\H)$),
one can define a $l$-grading by declaring $A$ of
$l$-degree $r=(r_1,\cdots,r_l)\in\beta\Z^l$ when
$$
V_zAV_{-z}=e^{-i(r_1z_1+\cdots+r_lz_l)}A,\;\forall z\in\T^l_\beta.
$$
Then, any $\a$-norm smooth operator can
be uniquely written as a norm convergent sum
$$
A=\sum_{r\in\beta\Z^l}A_r,
$$
where each $A_r$ is of $l$-degree $(r_1,\cdots,r_l)$.\\
This is in
particular the case for the operator of pointwise multiplication by
any function  $f\in \Coo_c(M)$, since
$M_f$ lies inside the smooth domains of the derivations
$\delta_j(.):=[X_j,.]$. This assertion is obtained iterating the relation
$$
\|[X_j,M_f]\|=\|M_{X_j(f)}\|=\|X_j(f)\|_\infty,
$$
which is finite since $f\in\Coo_c(M)$ and because the $X_j$ are ordinary
smooth vector fields.\\
Writing the spectral subspace decomposition of such operator, we find
the Peter-Weyl decomposition of any $f\in C^\infty(M)\cap
L^\infty(M,\mu_g)$, as $f=\sum_{r\in\beta\Z^l}f_r$, where $f_r$
satisfies $\a_z(f_r)=e^{-i(r_1z_1+\cdots+r_lz_l)}f_r$. The twisted
product of homogeneous components satisfies the noncommutative
torus relation:
\begin{equation}
\label{NTR}
f_r\Mop h_s=e^{-\frac{i}{2} <r,\Th s>}f_r\,h_s.
\end{equation}
Noncommutative tori $\T^n_\Th$, odd and even Connes--Landi spheres
$\Sf^{2n+1}_\th,\Sf^{2n}_\th$~\cite{ConnesDV} are examples of such
compact noncommutative spaces; and the ambient space of
$\Sf^{n-1}_\th$ is a non-compact periodic deformation.

In summary, it is clear that the noncommutative structures of
isospectral deformations are inherited from the NC tori or Moyal
planes one's, depending whether the deformation is periodic or not.

When $\Th$ is not invertible, the deformed product reduces to another twisted
product associated with the restricted action $\sigma:=\a|_{V^\bot}$, where
$V$ is the null space of $\Th$ ---see for example \cite{RieffelDefQ}. Hence,
one can handle non-invertible deformation matrices without any trouble. But of
course, the ``effective'' deformation is always of even rank.

Finally, in the non-periodic case only, properness of~$\a$ implies that it is
also free. To see that, recall that properness of any $G$-action is equivalent
to $\{g\in G|g.X\cap Y\ne\emptyset\}$ is compact for any $X,Y$ compact subset
of $M$ ---see \cite{Michor}. So, taking $X=Y=\{p_0\}$ for any $p_0\in M$, its
isotropy group $H_{p_0}=\{z\in\R^l|z.p_0=p_0\}=\{z\in\R^l|z.\{p_0\}
\cap\{p_0\}\ne\emptyset\}$ is compact as well. But the only compact subgroup of
$\R^l$ is $\{0\}$, hence the action is automatically free. This implies that
the quotient map $\pi:M\to M/\R^l$ defines a $\R^l$-principal bundle projection.

In the periodic case, the action is no longer automatically free, and
the set $M_{sing}$ of points with non-trivial isotropy groups can give
rise to additional divergences in the effective action. This will be
shown to constitute a new feature of the UV/IR mixing on isospectral
deformation manifolds.

\section{$\vf^{\Mop 4}$ theory on 4-d isospectral deformations}
\subsection{The effective action at one-loop}

For the sake of simplicity, we now restrict to the four-dimensional
case; $n=\dim(M)=4$. It will be clear, nevertheless, that our techniques apply to
higher dimensions without essential modifications. We consider the
classical functional action for a real scalar field $\vf$:
\begin{equation}
\label{eq:actioncl}
S[\vf] := \int_M\,\mu_g\,\Bigl[\thalf(\nabla^\mu\vf)\Mop
(\nabla_\mu\vf) +\thalf m^2\vf\Mop\vf+\frac{\lambda}{4!}\vf^{\Mop
4}\Bigr].
\end{equation}
We could add a coupling with gravitation of the type $\xi
R(\vf\Mop\vf)$ (or even $\xi R\Mop\vf\Mop\vf$),
where $R$ is the scalar curvature and $\xi$ a
coupling constant, without change in our conclusions. Indeed,
this term is not modified by the deformation:
due to the $\a$-invariance of the scalar curvature, we have
$R\Mop f=R.f$ for any $f\in\Coo_c(M)$, thus
$$
\int_M\mu_g\,R.(\vf\Mop\vf)=
\int_M\mu_g\,R\Mop\vf\Mop\vf=
\int_M\mu_g\,(R\Mop\vf).\vf=
\int_M\mu_g\,R\,.\vf\,.\vf.
$$
Similarly, thanks to the trace
property~(\ref{eq:trace}), $S[\vf]$ can be rewritten as
\begin{equation}
S[\vf] = \int_M\,\mu_g\,\Bigl[\thalf\vf\tri\vf\
+ \thalf m^2\vf\,\vf+\frac{\lambda}{4!}(\vf\Mop\vf)\,(\vf\Mop\vf)\Bigr],
\end{equation}
so that, as in the falt cases, the kinetic part is not affected by the deformation.
Recall that in our conventions the Laplacian is positive:
$\tri=-\nabla^\mu\nabla_\mu$.

\smallskip

We aim to compute the divergent part of the effective
action~$\Ga_{1l}[\vf]$ associated to~$S[\vf]$ at one loop. This is
formally given by $\thalf \ln(\det H) $, where $H$ is the effective
potential. In our case (as in the commutative one) it will be seen
that $H=\tri+m^2+B$, where $B$ is positive and bounded; so that when
the manifold is not compact $H$ has a non empty essential spectrum
(typically the whole interval $[m^2,+\infty[$). In order to deal with
operators having pure-point spectrum (discret with finite multiplicity),
we need first (independently of any regularization scheme) to
redefine formally the one-loop effective action as:
$$
\Ga_{1l}[\vf] := \thalf\ln\det\big(HH_0^{-1}\big),
$$
where $H_0^{-1}:=(\tri+m^2)^{-1}$ is the free propagator. We are ``not
so far'' from having a well defined determinant since:
$$
HH_0^{-1}=(H_0+B)H_0^{-1}=1+BH_0^{-1},
$$
and $BH_0^{-1}$ is `small': not trace-class in general, but compact;
more precisely $BH_0^{-1}$ lies inside the $p$-th Schatten-class for
all $p>2$ (see below for the concrete expression of $B$ and \cite{GIV}
for a proof of this claim). Physically, to replace $H$ by $HH_0^{-1}$
corresponds to remove the vacuum-to-vacuum amplitudes. We then define
the logarithm of the determinant by the Schwinger ``proper time''
representation:
\begin{equation}
\label{eq:actioneff}
\Ga_{1l}[\vf] = \frac{1}{2}\ln\big(\det(HH_0^{-1})\big)
:=-\frac{1}{2}\int_0^\infty\frac{dt}{t}\;\Tr\left(e^{-tH}-e^{-tH_0}\right).
\end{equation}
Before giving a precise meaning to the previous expression, that is to choose
a regularization scheme, we go through the computation of the effective
potential~$H$. For that, the following definition will be useful.
\begin{defn}
Let $(X,d\mu)$ a measure space. A kernel operator on $\mathcal{E}$, a functions space
on $X$, is a linear map $A:\mathcal{E}\to\mathcal{E}$ which can be written as
$$
\big(Af\big)(p)=\int_X\,d\mu(q)\,K_A(p,q)\,f(q),\hspace{0.5cm}f\in\mathcal{E},\,\,
p,q\in X,
$$
where $K_A$ is the kernel of $A$. This definition leads to the following rules
for the product of two kernel operators and for the kernel of the adjoint:
\begin{equation}
\label{kerprod}
K_{AB}(p,q)=\int_X\,d\mu(u)\,K_A(p,u)\,K_B(u,q),\sepword{and}
K_{A^*}(p,q)=K_A(q,p)^*.
\end{equation}
\end{defn}
In our case, $(X,d\mu)\equiv (M,\mu_g)$ as a measure space,
$\mathcal{E}\equiv \Coo_c(M)$ and we will only be interested
on distributional kernels, that is those $K_A$ lying on
$\Coo_c(M\x M)'$, the space of distributions on $M\x M$.

Recall that the effective potential (see for example
\cite{ZJ}) is the operator whose distributional kernel is given by the second
functional derivative of the classical action:
$$
K_H(p,p') := \frac{\delta^2S[\vf]}{\delta\vf(p)\delta\vf(p')},\;
\hspace{1cm} K_{H_0}(p,p') := \left.\frac{\delta^2S[\vf]}
{\delta\vf(p)\delta\vf(p')}\right|_{\lambda=0},
$$
with functional derivatives  defined as usual in the weak sense
$$
\left\langle\frac{\delta S[\vf]}{\delta\vf},\psi\right\rangle:=
\left.\frac{d S[\vf+t\psi]}{dt}\right|_{t=0},
$$
where the coupling is given by the integral with Riemannian volume
form $\left\langle f,h\right\rangle=\int_M\,\mu_g \,f\,h$.

Using the trace property~(\ref{eq:trace}) we find out:
\begin{equation*}
\left.\frac{d S[\vf+t\psi]}{dt}\right|_{t=0}=
\left\langle \tri\vf+m^2\vf+\frac{\lambda}{3!}\vf^{\Mop 3},\psi\right\rangle.
\end{equation*}
Hence,
$$
\tilde{S}_p[\vf]:=\frac{\delta S[\vf]}{\delta\vf(p)}
=\tri\vf(p)+m^2\vf(p)+\frac{\lambda}{3!}\vf^{\Mop 3}(p).
$$
The second functional derivative reads
\begin{align*}
\left\langle\frac{\delta^2S[\vf]}{\delta\vf(p)\delta\vf} ,\psi\right\rangle
:=&\left.\frac{d \tilde{S}_p[\vf+t\psi]}{dt}\right|_{t=0}\\
=&\left\langle \big(\tri+m^2+\frac{\lambda}{3!}(L_{\vf\Mop\vf}+R_{\vf\Mop\vf}
+R_{\vf}L_{\vf})\big)\delta_p^g,\psi\right\rangle,
\end{align*}
where $\delta_p^g$ is the distribution defined by $\langle \delta_q^g,\phi\rangle=
\int_M\mu_g(p)\delta_q^g(p)\phi(p)=\phi(q)$, for any test function $\phi\in
C^\infty_c(M)$.\\
In conclusion, the explicit form of the operator $H$ is:
$$
H = \tri+m^2+\frac{\lambda}{3!}(L_{\vf\Mop\vf}+R_{\vf\Mop\vf}
+R_{\vf}L_{\vf}).
$$
Because $\vf$ is real, the operators $L_\vf$ and $R_\vf$ are
self-adjoint, and we can check directly the strict positivity of~$H$:
$$
L_{\vf\Mop\vf}+R_{\vf\Mop\vf}+L_\vf R_\vf=
\thalf(L_\vf+R_\vf)^*(L_\vf+R_\vf)+\thalf L_\vf^*L_\vf+\thalf R_\vf^*R_\vf.
$$

\smallskip

We are come to an important point: the existence of UV/IR mixing for field
theory on isospectral deformations comes from the simultaneous presence of
left and right twisted multiplications in the effective potential. Precisely,
we wish to illustrate the smearing nature of the product of left and right
twisted multiplication operator $L_fR_h$. The crucial consequence, employed in
subsection 3.3, is that the trace of $L_f\,R_h\,e^{-t(\tri+m^2)}$ is regular
when $t$ goes to zero, contrary to $\Tr(L_f\,e^{-t(\tri+m^2)}),
\Tr(R_f\,e^{-t(\tri+m^2)}),\Tr(M_f\,e^{-t(\tri+m^2)})$, which in
$n$~dimensions behave as~$t^{-n/2}$ when $t\to 0$ (In fact the three latter
traces are identical).
\begin{rem}
For a $\frac{\lambda}{3!}\vf^{\Mop 3}$ theory on a six dimensional manifold,
the effective potential reads:
$$
H=\tri+m^2+\frac{\lambda}{2!}(L_\vf+R_\vf).
$$
Even in the lack of the `mixed' term $R_\vf L_\vf$, those theories have a non-planar
sector, but which will be present only at the level of the two-point function;
the tadpole is not affected by the mixing.
\end{rem}
Consider the non-degenerate ($n=2N,\Th$ invertible) Moyal plane case. The
operator $L_fR_h$ turns out to be trace-class whenever
$f,h\in\SS(\R^{2N})$, say. This fact is known to the experts, but rarely
mentioned ---to the knowledge of the author, its first mention in writing is
in~\cite{Braunss}. We do a little disgression to see how it comes about.
Recall~\cite{Himalia} that there is an orthonormal basis for
$L^2(\R^{2N},d^{2N}x)$, the harmonic oscillator eigentransitions
$(2\pi\th)^{-N/2}\{f_{mn}\}_{m,n\in\N^N},\th:=(\det\Th)^{1/2N}$ which are
matrix units for the Moyal product:
$$
f_{mn}\Mop f_{kl}=\delta_{nk}f_{ml}.
$$
Expanding $f,h\in\SS(\R^{2N})$ in this basis:
$f=\sum_{m,n}c_{mn}f_{mn},h=\sum_{m,n}d_{mn}f_{mn}$, we obtain:
\begin{align*}
\Tr\big(L_fR_h\big)=&(2\pi\th)^{-N}\sum_{m,n,k,l,s,t}c_{kl}\,d_{st}\,
\langle f_{mn},f_{kl}\Mop f_{mn}\Mop f_{st}\rangle\\
=&(2\pi\th)^{-N}\sum_{m,n,k,t}c_{km}\,d_{nt}\,
\langle f_{mn},f_{kt}\rangle\\
=&\sum_{m,n}c_{mm}\,d_{nn}\\
=&(2\pi\th)^{-N}\int d^{2N}x\,f(x) \int d^{2N}y\,h(y) < \infty.
\end{align*}
Then in this case one can factorize $HH_0^{-1}$ and extract a finite
part in the effective action. We have
\begin{align}
\label{bel}
H_0H^{-1}=&
\Big(1-\frac{\lambda}{3!}(L_{\vf\mop\vf}+R_{\vf\mop\vf})
\frac{1}{\tri+m^2 +\frac{\lambda}{3!}(L_{\vf\mop\vf}+R_{\vf\mop\vf})}\Big)\nonumber\\
&\hspace{1cm}\times\Big(1-\frac{\lambda}{3!}L_\vf R_\vf
\frac{1}{\tri+m^2
+\frac{\lambda}{3!}(L_{\vf\mop\vf}+R_{\vf\mop\vf}+L_\vf R_\vf)}\Big).
\end{align}
Now,
$$
1-\frac{\lambda}{3!}L_\vf R_\vf
\frac{1}{\tri+M^2
+\frac{\lambda}{3!}(L_{\vf\mop\vf}+R_{\vf\mop\vf}+L_\vf R_\vf)}
\in 1+\L^1(\H),
$$
so that its determinant is well-defined. Thus only the determinant of
the first piece of (\ref{bel}) needs to be regularized. The determinant of the second
piece of (\ref{bel}) contains the whole non-planar contribution to the two-point
function, while for the four-point function the finite non-planar part
lies in both pieces.

The structure of the effective potential, i.e. the presence of mixed products
of left and right twisted multiplication operators, and thus the existence of
two distinct sectors in the theory is fairly general: for noncommutative scalar field
theories whose classical field counterparts are regarded as elements of a
noncommutative algebra, and a classical action built from a trace on the
algebra, the effective potential will contain in general sums and mixed
products of left and right regular representation operators.

Let us go back to the computation of~$\Ga_{1l}[\vf]$. The $t$-integral
in (\ref{eq:actioneff}) is divergent because of the small-$t$ behavior
of the heat kernel on the diagonal. We thus define a one-loop
regularized effective action by:
\begin{equation}
\label{eq:actioneffreg}
\Ga^\ep_{1l}[\vf]:=-\frac{1}{2}\int_\ep^\infty\frac{dt}{t}\;
\Tr\left(e^{-tH}-e^{-tH_0}\right).
\end{equation}
One can invoke less rough regularization schemes, for example a
$\zeta$-function regularization
\begin{equation}
\label{eq:p}
\Ga^{\sigma,\mu}_{1l}[\vf]:=-\frac{1}{2}\int_0^\infty\frac{dt}{t}
\left(t\mu^2\right)^\sigma\;\Tr\left(e^{-tH}-e^{-tH_0}\right),
\end{equation}
akin to dimensional regularization. However, for the purposes of this
article (\ref{eq:actioneffreg}) will do. One can think of $\epsilon$
as of the inverse square of $\Lambda$, with~$\Lambda$ a momentum space
cutoff.

To show that the expressions~(\ref{eq:actioneffreg}) and~(\ref{eq:p})
are now well defined, we have to prove that $e^{-tH}-e^{-tH_0}$ is
trace-class for all $t>0$. Note that for $t\to\infty$ convergence
is ensured by the global $e^{-tm^2}$ factor, and that when the
spectrum of the Laplacian has a strictly positive lower bound one can
construct massless, IR~divergence-free NCQFT. That is the case for the
twisted hyperbolic planes $\HH^n_\Th$ since the $L^2$-spectrum of
$\tri$ on $\HH^n$ is the whole half line $[n^2/4,\infty[$.
\begin{lem}
The semigroup difference $e^{-tH}-e^{-tH_0}$ is trace-class for all
$t>0$.
\end{lem}
\begin{proof}
Using positivity of $H$ and $H_0$, the semigroup property and the
holomorphic functional calculus with a path $\ga$ surrounding both the
spectrum $\spec(H)\subset\R^+$ and $\spec(H_0)\subset\R^+$, we have
$$
e^{-tH}-e^{-tH_0}=\frac{1}{(2i\pi)^2}
\int_{\ga\times\ga}dz_1\,dz_2\,e^{-t(z_1+z_2)/2}\,
\left(R_H(z_1)R_H(z_2)-R_{H_0}(z_1)R_{H_0}(z_2)\right),
$$
where $R_A(z)=(z-A)^{-1}$ denotes the resolvent of $A$. But $H=H_0+B$ where $B$ is
bounded. Using next $R_H(z)=R_{H_0}(z)(1+BR_H(z))$, we find
\begin{align*}
R_H(z_1)R_H(z_2)-R_{H_0}(z_1)R_{H_0}(z_2)=&
R_{H_0}(z_1)R_{H_0}(z_2)BR_H(z_2)\\
&+R_{H_0}(z_1)BR_H(z_1)R_{H_0}(z_2)\\
&+R_{H_0}(z_1)BR_H(z_1)R_{H_0}(z_2)BR_H(z_2).
\end{align*}
The first resolvent equation and the fact that
$L_f(z-\tri)^{-k},R_f(z-\tri)^{-k}\in\L^p(\H)$, for
$p>2/k,f\in\Coo_c(M)$~\cite{GIV}, together with the H\"older inequality
for Schatten classes, yield
$$
\int_{\ga\times\ga}dz_1\,dz_2\,e^{-t(z_1+z_2)/2}
R_{H_0}(z_1)R_{H_0}(z_2)BR_H(z_2)
$$
is absolutly convergent for the trace norm. Similarly for the other terms. So
$e^{-tH}-e^{-tH_0}$ is trace-class as required.
\end{proof}

\subsection{Field expansion}

We now tackle the $\ep$-behavior of $\Ga^\ep_{1l}[\vf]$ to describe
the divergences. We will then show that, as for the Moyal planes and
noncommutative tori, there exist for general isospectral deformations
two kind of contributions to the Green functions, the planar one
giving rise to ordinary singularities and the non-planar one
exhibiting the UV/IR mixing phenomenon. Note that, since we are in a
curved background, we can no longer work with Feynman diagrams in
momentum space. However, by abuse of language we continue to speak
about planar and non-planar contributions, because there is a
splitting at the operator level which coincides with the splitting of
planar and non-planar Feynman graphs in the known flat cases. This
point will become clearer in subsequent subsections.

As we are only interested in the $\ep$-behavior of $\Ga^\ep_{1l}[\vf]$
(we only consider the potentially divergent part of the regularized
effective action), we need a small $t$-expansion for $\Tr\left(e^{-tH}-
e^{-tH_0}\right)$. This expansion will be managed in the same vein as the ones
obtained in \cite{GI,Vass}. The Baker-Campbell-Hausdorff formula is written:
\begin{equation}
\label{poiu}
e^{-tH}=e^{-tB+\frac{t^2}{2}[\tri ,B]
-\frac{t^3}{6}[\tri ,[\tri ,B]]-\frac{t^3}{12}[B,[\tri ,B]]+\cdots}\;
e^{-tH_0}.
\end{equation}
We now expand the first exponential up to factors which, after taking
the trace, give terms of order less or equal to zero in $t$. Only a
few terms will be important: \\
We have first to take into account that (in $n$
dimensions)
\begin{align}
\Tr(L_f\tri^k e^{-t\tri})\simeq& t^{-n/2-k},\,t\rightarrow 0,\nonumber\\
\Tr(R_f\tri^k e^{-t\tri})\simeq&t^{-n/2-k},\,t\rightarrow 0.
\label{eq:dhk}
\end{align}
Indeed, for the ``left'' case (the right one being similar) since, as proved
in~\cite{GIV}, one has $L_f(1+\tri)^{-k}\in\L^p(\H)$ for any $p>n/2k$
and any $f\in\Coo_c(M)$, we conclude for all $\ep>0$:
\begin{align*}
\|L_f\tri^ke^{-t\tri}\|_1\leq&\|L_f(1+\tri)^{-n/2-\ep}\|_1\,\,
\|\frac{\tri^k}{(1+\tri)^k}\|\|(1+\tri)^{n/2+k+\ep}e^{-t\tri}\|\\
\leq&C(\ep)t^{-(n/2+k+\ep)}.
\end{align*}
The last estimate follows from functional calculus.
Therefore, in the
field expansion we need to correct the power in $t$ by the order of
the differential operator appearing when we expand the first exponential
in the equation \eqref{poiu}.\\
Secondly, we have to notice that the commutators $[\tri,L_f]$, $[\tri,R_f]$
(and also $[\tri,R_fL_h]$) reduce by one the order of the differential
operator (cf. equation \eqref{pour} below).
To see this, we compute the commutators $[\tri,L_f]$, $[\tri,R_f]$ and $[\tri,R_fL_h]$.
The simplest way is to use the formulas (\ref{Lfint}) and
(\ref{Rfint}). By $[V_z,\tri]=0$ for all $z\in\R^l$ (from the isometry
property of $\a$) and choosing a local coordinate system $\{x^\mu\}$,
one obtains
\begin{align}
\label{pour}
[\tri,L_f]&=
(2\pi)^{-l}\int_{\R^{2l}}\,d^ly\,d^lz\;e^{-i<y,z>}\;
V_{\thalf \Th y}\,[\tri,M_f]\,V_{-\thalf\Th y-z}\nonumber\\
&=
(2\pi)^{-l}\int_{\R^{2l}}\,d^ly\,d^lz\;e^{-i<y,z>}\;
V_{\thalf \Th y}\,\left(M_{\tri f}
-2M_{\nabla^\mu f}\nabla_\mu\right)\,V_{-\thalf\Th y-z}\nonumber\\
&=L_{\tri f}-2L_{\nabla^\mu f}\nabla_\mu,
\end{align}
and similarly,
\begin{align}
\label{pour2}
[\tri,R_f]&=R_{\tri f}-2R_{\nabla^\mu f }\nabla_\mu,\\
[\tri,R_fL_h]&=R_f[\tri,L_h]+[\tri,R_f]L_h\nonumber\\
&=R_fL_{\tri(h)}+R_{\tri(f)}L_h-2R_{\nabla^\mu f}L_{\nabla_\mu h}
-2\big(R_fL_{\nabla^\mu h}+R_{\nabla^\mu f}L_h\big)\nabla_\mu.
\end{align}
The local coordinate system used must be compatible with the
deformation, that is, defined on some $\a$-invariant open neighborhood
$U\subset M$. To obtain one such, choose any open covering
$\{U_I\}_{i\in I}$ of $M$ and define $\{\widetilde{U}_I\}_{i\in I}$ by
letting $\R^l$ act on it: $\widetilde{U}_i:=\R^l.U_i$.

\smallskip

This implies that in $n$ dimensions, one only
needs to use the BCH formula up to order $n-2$ to capture the
divergent structure of the effective action.

\smallskip

Moreover, that the commutators decrease the
degree of the differential operator is a necessary condition to make the BCH expansion
meaningfull: In \cite{Melpomene}, we considere a field theory
on a noncommutative 4-plane with an (associative)
position-dependant Moyal product (coming from a rank-2
Poisson structure on $\R^4$). It turns out that
the commutators $[\tri,L_f]$ and $[\tri,R_f]$ contain now a term with
an order two differential operator. This makes the BCH
development useless since the $k$-times iterated commutator
$[t\tri,[\cdots,[t\tri,t L_f]\cdots]]$ contains a term which gives
after the exponential expansion a contribution of order $t^{-n/2+1}$,
independently of $k$, the number of commutators involved. Thus, in this case the
whole BCH serie will be needed to capture the divergences.

\medskip

Putting all together, we finally obtain:
$$
e^{-tH}=\left(1-tB+\frac{t^2}{2}[\tri ,B]
-\frac{t^3}{6}[\tri ,[\tri ,B]]+\frac{t^2}{2}B^2\right)
e^{-tH_0}\;+O(t);
$$
we mean by this estimate that we have a small-$t$ expansion:
\begin{equation}
\label{rout}
\Tr\left(e^{-tH}-e^{-tH_0}\right)=
\Tr\left(\big(-tB+\frac{t^2}{2}[\tri,B]
-\frac{t^3}{6}[\tri,[\tri,B]]+\frac{t^2}{2}B^2\big)
e^{-tH_0}\right)\;+O(t).
\end{equation}

\smallskip

We now show that in fact, the commutators in the expression
\eqref{rout} give no contribution to the effective action. Indeed,
if each terms $C\tri e^{-t\tri}$ and $\tri C e^{-t\tri}$ are trace-class, with $C=B$
or $C=[\tri,B]$, then by the cyclicity of the trace and the fact that the Laplacian
commutes with the heat semigroup, one gets
\begin{equation}
\label{plk}
\Tr\big(\tri\, C \,e^{-t\tri}-C\,\tri\, e^{-t\tri}\big)
=\Tr\big(C\,\tri \,e^{-t\tri}-C\,\tri \,e^{-t\tri}\big)=0
\end{equation}
That $C\tri e^{-t\tri}$ is trace-class is obvious from fonctional calculus
and using the same arguments than those used to obtain the estimate \eqref{eq:dhk}.
For $\tri C e^{-t\tri}$, it is a little bit less immediat since
the latter appears as a product of a trace-class operator ($C e^{-t\tri}$) times an unbounded
one ($\tri$). Actually, using the tautological relation
$$
\tri\, C\, e^{-t\tri}=C\,\tri\,e^{-t\tri}+[\tri, C]\, e^{-t\tri},
$$
and the equations \eqref{pour} and \eqref{pour2} (iterated once more when
$C=[\tri,B]$), one sees that this term appears also as a sum of trace-class
operators. Hence \eqref{plk} is proved and we are left with
\begin{align*}
\Tr\Big(e^{-tH}-e^{-tH_0}\Big)=&
-t\frac{\lambda}{3!}\Tr\Big(\big(L_{\vf\Mop\vf}
+R_{\vf\Mop\vf}+R_\vf L_\vf\big)e^{-t(\tri+m^2)}\Big)\\
&+\frac{t^2}{2}\frac{\lambda^2}{(3!)^2}\Tr\Big(\big(L_{\vf^{\Mop 4}}
+R_{\vf^{\Mop 4}}+3R_{\vf\Mop\vf}L_{\vf\Mop\vf}\\
&\hspace{3cm}+2R_\vf L_{\vf^{\Mop 3}}
+2R_{\vf^{\Mop 3}}L_\vf\big)e^{-t(\tri+m^2)}\Big)+O(t).
\end{align*}

\subsection{Planar and non-planar contributions}

We split the previous expansion in two parts. In the first one, we
only keep terms like $L_fe^{-t\tri}$ and $R_fe^{-t\tri}$. Those belong to
the ``planar part'', since they give commutative-like contributions as
easily seen from equation~(\ref{eq:deq}) below. The second
contribution, corresponding to the ``non-planar part'', consists of
crossed terms like $L_f R_he^{-t\tri}$.

The planar contribution to the effective action is
\begin{align*}
\Ga_{1l,P}^\ep[\vf]:=&\frac{1}{2}\int_\ep^\infty dt\,e^{-tm^2}\Big\lbrace
\frac{\lambda}{3!}\Tr\Big(\big(L_{\vf\Mop\vf}
+R_{\vf\Mop\vf}\big)e^{-t\tri}\Big)\\
&\hspace{3cm}-\frac{t}{2}\frac{\lambda^2}{(3!)^2}
\Tr\Big(\big(L_{\vf^{\Mop 4}}
+R_{\vf^{\Mop 4}}\big)e^{-t\tri}\Big)\Big\rbrace+O(\ep^0).
\end{align*}
To compute those traces, let us show that first the trace is a
dequantizer for the deformed product
\begin{equation}
\label{eq:deq}
\Tr\big(L_f\,e^{-t\tri}\big)=\Tr\big(R_f\,e^{-t\tri}\big)=\Tr\big(M_f\,e^{-t\tri}\big),
\end{equation}
whenever $M_f\,e^{-t\tri}$ is trace-class. Here $M_f$ still denotes the
operator of pointwise multiplication by $f$. We only treat the $L_f$
case, since for the $R_f$ case the arguments are similar. From the
definition \ref{mop} and the product rule (\ref{kerprod}) for kernel operators, a little
calculation gives the following expression for the Schwartz kernel of
$L_f\,e^{-t\tri}$:
$$
K_{L_f\,e^{-t\tri}}(p,p')=(2\pi)^{-l}\int_{\R^{2l}}\,d^ly\,d^lz\,e^{-i<y,z>}\,
f(-\thalf\Th y.p)\,K_t(z.p,p').
$$
Then
\begin{align*}
\Tr\big(L_f\,e^{-t\tri}\big)&=\int_M\,\mu_g(p)\,K_{L_f\,e^{-t\tri}}(p,p)\\
&=(2\pi)^{-l}\int_M\,\mu_g(p)\int_{\R^{2l}}\,d^ly\,d^lz\,e^{-i<y,z>}\,
f(-\thalf\Th y.p)\,K_t(z.p,p).
\end{align*}
Using next the invariance of the volume form under the isometry
$p\rightarrow \thalf\Th y.p$ and the fact that $[e^{-t\tri},V_z]=0$,
translated in terms of invariance of its kernel
\begin{equation}
\label{invprop}
K_t(z.p,z.p')=K_t(p,p'),
\end{equation}
the claim follows after a plane waves integration:
\begin{align*}
\Tr\big(L_f\,e^{-t\tri}\big)
&=(2\pi)^{-l}\int_M\,\mu_g(p)\int_{\R^{2l}}\,d^ly\,d^lz\,e^{-i<y,z>}\,
f(p)\,K_t(z.p,p)\\
&=\int_M\,\mu_g(p)\,f(p)\int_{\R^{l}}\,d^lz\,\delta(z)\,
K_t(z.p,p)\\
&=\int_M\,\mu_g(p)\,f(p)\,K_t(p,p)=
\Tr\big(M_f\,e^{-t\tri}\big).
\end{align*}
Hence, the planar part of the one loop effective action reads:
\begin{equation*}
\Ga_{1l,P}^\ep[\vf]=\int_\ep^\infty dt\,e^{-tm^2}\Big\lbrace
\frac{\lambda}{3!}\Tr\Big(M_{\vf\Mop\vf}e^{-t\tri}\Big)
-\frac{t}{2}\frac{\lambda^2}{(3!)^2}\Tr\Big(M_{\vf^{\Mop 4}}
e^{-t\tri}\Big)\Big\rbrace + O(\ep^0).
\end{equation*}
Using the on-diagonal heat kernel expansion up to order one
$$
K_t(x,x)=(4\pi t)^{-2}\bigl(1-\frac{t}{6}R(x)\bigr)+O(t^0),
$$
where $R$ is the scalar curvature, together with the relation
$$
K_{M_fe^{-t\tri}}(x,x)=f(x)K_t(x,x),
$$
one obtains at $\ep^0$~order:
\begin{equation}
\label{eq:P}
\Ga_{1l,P}^\ep[\vf]=\int_\ep^\infty \frac{dt\,e^{-tm^2}}{(4\pi t)^2}
\int_M\,\mu_g\,\Big(\frac{\lambda}{3!}\vf\Mop\vf
-t\Big(\frac{1}{6}\frac{\lambda}{3!}(\vf\Mop\vf)R
+\frac{1}{2}\frac{\lambda^2}{(3!)^2}\vf^{\Mop 4}\Big)\Big).
\end{equation}
The planar part thus yields ordinary $\frac{1}{\ep}$ and $|\ln\ep|$
divergences. They can be substracted adding local counter-terms to
the original action.

\smallskip

The contribution for the non-planar part is
\begin{align*}
\Ga_{1l,NP}^\ep[\vf]:=&\frac{1}{2}\int_\ep^\infty dt\,e^{-tm^2}\Big\lbrace
\frac{\lambda}{3!}\Tr\Big(R_\vf L_\vf\,e^{-t\tri}\Big)\\
&\hspace{1cm}-\frac{t}{2}\frac{\lambda^2}{(3!)^2}
\Tr\Big(\big(3R_{\vf\Mop\vf}L_{\vf\Mop\vf}
+2R_\vf L_{\vf^{\Mop 3}}+2R_{\vf^{\Mop 3}}L_\vf\big)e^{-t\tri}\Big)\Big\rbrace
+O(\ep^0).
\end{align*}
We now simplify this expression. By the definition of the twisted
product (\ref{mop}) and using the identity
$\psi(z.p)=\int_M\mu_g(p')\,\delta^g_{z.p}(p')\,\psi(p')$, one can
easily derive the Schwartz kernel of the left and right twisted
multiplication operators:
$$
K_{L_f}(p,p')=(2\pi)^{-l}\int_{\R^{2l}}\,
d^ly\,d^lz\,e^{-i<y,z>}f(-\thalf\Th y.p)\,\delta^g_{z.p}(p'),
$$
and
$$
K_{R_f}(p,p')=(2\pi)^{-l}\int_{\R^{2l}}\,
d^ly\,d^lz\,e^{-i<y,z>}f(z.p)\,\delta^g_{-\thalf\Th y.p}(p').
$$
By the kernel composition rule (\ref{kerprod}), we obtain after few changes
of variables and a plane waves integration, the kernel of
$L_fR_h e^{-t\tri}$ in term of the heat kernel $K_t$:
$$
K_{L_fR_h e^{-t\tri}}(p,p')=(2\pi)^{-l}
\int_{\R^{2l}}\,d^ly\,d^lz\,e^{-i<y,z>}\,f((-\thalf\Th y-z).p)\,
h(z.p)\,K_t(-\thalf\Th y.p,p').
$$
Hence, the trace of $L_fR_h e^{-t\tri}$ reads (with a few changes of
variable):
\begin{align}
\label{eq:RL}
\Tr\big(L_fR_h e^{-t\tri}\big)&=(2\pi)^{-l}
\int_M\mu_g(p)\,\int_{\R^{2l}}\,d^ly\,d^lz\,e^{-i<y,z>}\,
f(p)\,h(z.p)\,K_t(-\Th y.p,p)\nonumber\\
&=\Tr\big(R_fL_h e^{-t\tri}\big).
\end{align}
To obtain the last equality, we used the fact that $K_t$ is
symmetric, its invariance under $\a$ and the isometry $p\mapsto-z.p$.
Invoking formula (\ref{eq:RL}), we obtain for~$\Ga_{1l,NP}^\ep[\vf]$:
\begin{align}
\label{bell}
\Ga_{1l,NP}^\ep[\vf]=&
(2\pi)^{-l}\frac{1}{2}\int_\ep^\infty dt\,e^{-tm^2}
\int_M\mu_g(p)\,\int_{\R^{2l}}\,d^ly\,d^lz\,e^{-i<y,z>}
\Big\lbrace\frac{\lambda}{3!}\vf(p)\vf(z.p)\nonumber\\
&-\frac{t}{2}\frac{\lambda^2}{(3!)^2}\Big(3\vf\Mop\vf(p)\vf\Mop\vf(z.p)
+4\vf(p)\vf^{\Mop 3}(z.p)
\Big)\Big\rbrace K_t(-\Th y.p,p)+O(\ep^0)\nonumber.
\end{align}
We shall see that the better
$\ep$-behavior of the non-planar part and the UR/IV entanglement
phenomenon come from the presence of the off-diagonal heat kernel in
the previous expression. Depanding on the precise geometric setup, the non-planar contributions
could still be divergent. In the unfavourable cases, the divergences are non-local as
shown is the next subsections. This makes the renormalisation problematic.

\section{Non-periodic deformations}
\subsection{NCQFT on the Moyal plane in configuration space}

When $M=\R^4$ with the flat metric, $l=4$ and $\R^4$ acting on itself
by translation, isospectral deformation gives $\R^4_\Th$. In this
case, the heat kernel is exactly given by
$$
K_t(x,y)=(4\pi t)^{-2}e^{-\frac{|x-y|^2}{4t}},
$$
so we can explicitly compute $\Ga^\ep_{1l,P}(\vf)$ and $\Ga^\ep_{1l,NP}(\vf)$.
For the planar part, we obtain from~(\ref{eq:P})
$$\Ga_{1l,P}^\ep[\vf]=\int_\ep^\infty\,dt\frac{e^{-tm^2}}{(4\pi t)^2}\,
\int_{\R^4}\,d^4x\,\Big(\frac{\lambda}{3!}\vf^2(x)
-\frac{t}{2}\frac{\lambda^2}{(3!)^2}(\vf\Mop\vf)^2(x)\Big)+O(\ep^0),
$$
that will give the ordinary $\ep^{-1}$ and $|\ln\ep|$ divergences for
the respectively planar two- and four-point functions.

The non-planar part is given by:
\begin{align*}
\Ga_{1l,NP}^\ep[\vf]=&(2\pi)^{-4}\int_\ep^\infty dt\,
\frac{e^{-tm^2}}{(4\pi t)^2}\,
\int_{\R^{12}}\,d^4x\,d^4y\,d^4z\,e^{-i<y,z>}\,
e^{-\frac{|\Th y|^2}{4t}}
\Big(\frac{1}{2}\frac{\lambda}{3!}\vf(x)\vf(x+z)\\
&\hspace{1.5cm}-\frac{\lambda^2}{(3!)^2}\frac{t}{4}
\big(3\vf\Mop\vf(x)\vf\Mop\vf(x+z)
+4\vf(x)\vf^{\Mop 3}(x+z)\big)\Big)
+O(\ep^0).
\end{align*}
The Gaussian $y$-integration can be performed to obtain:
\begin{align*}
\Ga_{1l,NP}^\ep[\vf]=&(2\pi\th)^{-4}\int_\ep^\infty dt\,e^{-tm^2}\,
\int_{\R^8}\,d^4x\,d^4z\,e^{-t|\Th^{-1}(z-x)|^2}\\
&\times\Big(
\frac{1}{2}\frac{\lambda}{3!}\vf(x)\vf(z)
-\frac{\lambda^2}{(3!)^2}\frac{t}{4}\big(3\vf\Mop\vf(x)\vf\Mop\vf(z)
+4\vf(x)\vf^{\Mop 3}(z)\big)\Big)+O(\ep^0),
\end{align*}
where $\th:=(\det\Th)^{1/4}$. Finally, the $t$-integration gives
\begin{align*}
\Ga_{1l,NP}^\ep[\vf]=&(2\pi\th)^{-4}
\int_{\R^8}\,d^4x\,d^4z\,
\frac{e^{-\ep(m^2+|\Th^{-1}(z-x)|^2)}}{m^2+|\Th^{-1}(z-x)|^2}\\
&\times\Big(
\frac{\lambda}{2.3!}\vf(x)\vf(z)
-\frac{\lambda^2}{(3!)^2}\frac{3\vf\Mop\vf(x)\vf\Mop\vf(z)
+4\vf(x)\vf^{\Mop 3}(z)}{4(m^2+|\Th^{-1}(z-x)|^2)}\Big)+O(\ep^0).
\end{align*}
This expression is regular when $\ep$ goes to zero ---we are now in
the full noncommutative picture.

  From the previous formula one reads off the associated (non-planar)
two- and four-point functions in configuration space in the limit
$\ep\rightarrow 0$:
\begin{align*}
G_{1l,NP}^2(x,y)=&(\pi\th)^{-4}\frac{\lambda}{96}
\frac{1}{m^2+|\Th^{-1}(x-y)|^2},\\
G_{1l,NP}^4(x,y,z,u)=&-(\pi\th)^{-8}
\frac{\lambda^2}{24}\Big(\frac{3}{2}\delta(x-y+z-u)
\,\int d^4v\,\frac{e^{2i<v,\Th^{-1}(u-z)>}}{(m^2+|\Th^{-1}(z-v-x)|^2)^2}\\
&\hspace{4cm}+\frac{e^{2i<x-y,\Th^{-1}(z-y)>}}
{(m^2+|\Th^{-1}(x-y+z-u)|^2)^2}\Big).
\end{align*}

We see that the UV/IR mixing in configuration space manifests itself
in the long-range behavior of the correlation functions. The slow
decreasing at infinity of the two- and four-point functions is
equivalent to a IR~singularity in momentum space, as shown by a
Fourier transform:
$$
\widehat{G^2}_{1l,NP}(\xi,\eta) \propto \frac{m}{\th|\xi|}K_1(m|\Th\xi|)\delta(\xi+\eta).
$$
Here $K_n(z)$ denotes the $n$-th modified Bessel function. We retrieve
the known UV/IR mixing (se for example \cite{FRR}):
$$
\frac{m}{\th|\xi|}K_1(m|\Th\xi|)\sim(\th|\xi|)^{-2},\;|\xi|\to 0.
$$
This last result at one loop in the Moyal (translation-invariant)
context is usually obtained by means of Feynman diagrams in momentum
space ---see for example~\cite{FRR}. We just checked that the Fourier
transform for the two-point function coincides with the standard
calculation's result.

However, this is not the end of the story. The behavior of the
amplitudes as $\th\downarrow0$ presents interesting differences
in configuration and momentum spaces. Assume
that $\Th$ has been put in the canonical form
$$
\Th = \begin{pmatrix} & \th & & \\ -\th & & & \\ & & & \th' \\
& & -\th' & \end{pmatrix},
$$
and choose $\th'=\th$ for simplicity. In effect, developing the
two-point expression in terms of~$\th$, we find
$$
\frac{1}{\th^4 m^2 + \th^2|x|^2} = \frac{1}{\th^2|x|^2} \biggl(1 -
\frac{\th^2m^2}{|x|^2} + \frac{\th^4m^4}{|x|^4} - \cdots \biggr).
$$
First of all, we remark that the logarithmic dependence on $\th$ of
the UV/IR mixing in momentum space (in addition to its quadratic
divergence) found in~\cite{FRR} is apparently absent here.
Now, with the sole exception of the first term, the previous series is
made of functions that are not tempered distributions, and so they
have no Fourier transform. In other words, the passage to the
``commutative limit'' does not commute with taking Fourier transforms.

The question is subtler, though. We can ask ourselves to which kind of
divergences the terms of the last development are associated to. The
answer is that first term is infrared divergent in configuration
space; the second one is both ultraviolet and infrared divergent, and
the following are all ultraviolet divergent. It is perhaps surprising
that there is a way to recover the exact result from that nearly
nonsensical infinite series; this involves precisely the correction to
the indicated UV divergences. Indeed we can ``renormalize'' (in the
sense of Epstein and Glaser) the $1/|x|^{2k+4}$ functions, with the
result that the redefined distributions $[1/|x|^{2k+4}]_R$ are
tempered. Those $[.]_R$ distributions depend on a mass scale
parameter. Their Fourier transforms $\widehat{[1/|x|^{2k+4}]_R}$
(making a long history short) have been calculated as well~\cite{Carme,
NR}, with the result
$$
\widehat{[1/|x|^{2k+4}]_R}(\xi) =
\frac{(-)^{k+1}|\xi|^{2k}}{4^{k+1}k!(k+1)!}
\biggl[2\ln\frac{|\xi|}{2\mu} - \Psi(k+1) - \Psi(k+2)\biggr],
$$
Now, a natural mass scale parameter in our context is $1/\th m$. This
is where $\ln\th$ can sneak back in. Upon substituting this for $\mu$
in the previous formula, and summing the series of Fourier transforms,
we recover on the nose the exact result:
$$
\frac{1}{\theta^2|\xi|^2}+
\frac{m^2}{2} \sum_{n=0}^\infty \frac{\th^{2n}m^{2n}|\xi|^{2n}}{4^n\,n!(n+1)!}
\biggl(\ln\frac{\th m|\xi|}{2} - \Psi(n+1) - \Psi(n+2)\biggr) =
\frac{m}{\th|\xi|}K_1(\th m|\xi|).
$$
For the four-point function, again in the $\th\downarrow0$ limit no
dependence on~$\ln(\th)$ is apparent in configuration space. The resulting
expression is however (UV- and) IR-divergent, and its redefinition
\emph{\`a la} Epstein and Glaser allows one to reintroduce the $\ln\th$.

\smallskip

The effect of the rank of $\Th$ becomes clearer in position space.
Indeed, for a generic $n$-dimensional Moyal plane with a deformation
matrix of rank $l\leq n$, the two-point function in momentum space is
always finite and behaves as $|\Th\xi|^{-n+2}$, when $\xi\to 0$.
However, since $\Th\xi\in\im(\Th)=\R^l$, the IR~singularity is not
locally integrable if $l\leq n-2$. It follows that the
two-point Green function does not have a Fourier transform since it is not a
temperate distribution. Thus in the four-dimensional case,
the non-planar contribution to the tadpole in
position space remains infinite if $l=2$! The four-point function has a
Fourier transform, its IR~singularity in momentum space being of the $\ln$
type, and the Green function in position space is finite whenever $l\ne 0$.
For example, had we treated $\R^2_\th\times\R^2$ instead of $\R^4_\Th$, we
would have found that the four-point part of~$\Ga_{1l,NP}^\ep[\vf]$ is
convergent, while the two-point part diverges as $\ln\ep$. \\
This point is discussed in details in~\cite{Parthenope}, where we use
the $\zeta$-regularization scheme and the Duhamel asymptotic expansion (instead
of the BCH one), in order to compare our results with those present in the literature.

These features of
the UV/IR mixing phenomenon on position space reappear in the general
non-periodic case, where the effective action will still be divergent for
$l=2$. This is shown in the next subsection.

\subsection{The divergences of the general non-periodic case}

Assume $\vf\in\Coo_c(M)$. We have also to make some more precise
assumptions on the behavior of the geometry at infinity in order to
control the heat kernel. In \cite{Chavel, Davies}, it is proved that
if $M$ is non-compact, complete, with Ricci curvature bounded from
below (plus either uniform
boundness of the inverse of the volume or of the inverse of the
isoperimetric constant of the Riemannian ball for some fixed radius),
then the heat kernel satisfies
\begin{equation}
\label{heatkernelestimate}
(4\pi t)^{-2}e^{-d_g^2(p,p')/4t}\leq K_t(p,p')
\leq C(4\pi t)^{-2}e^{-d_g^2(p,p')/4(1+c)t},
\end{equation}
where $d_g$ is the Riemannian distance and $C,c$ are strictly
positive constants.

In the general periodic case, we have shown that $\Ga^{\ep}_{1l,NP}[\vf]$ is given by:
\begin{align*}
\Ga_{1l,NP}^\ep[\vf]=&\frac{1}{2(2\pi)^l}\int_\ep^\infty dt\,e^{-tm^2}
\int_M\mu_g(p)\,\int_{\R^{2l}}\,d^ly\,d^lz\,
e^{-i<y,z>}\,K_t(-\Th y.p,p)\\
&\times\Big\lbrace\frac{\lambda}{3!}\vf(p)\vf(z.p)
-\frac{t}{2}\frac{\lambda^2}{(3!)^2}\Big(3\vf\Mop\vf(p)\vf\Mop\vf(z.p)
+4\vf(p)\vf^{\Mop 3}(z.p)
\Big)\Big\rbrace+O(\ep^0).
\end{align*}
We now show that this expression cannot produce more important
divergences than the planar contribution. Again, the regularity of
those integrals depends only on $l$ (that we may call the effective
noncommutative dimension), and on the metric through the Riemannian
distance function.

Before estimating the two-point part of $\Ga_{1l,NP}^\ep[\vf]$, which
is our main purpose in this section, we make the following remark: in
our present setting, the two-point non-planar Green function reads
$$
G^\ep_{1l,NP,2P}(p,p')=\frac{\lambda}{6(2\pi)^l}\int_{\R^{2l}}d^ly\,d^lz\,
e^{-i<y,z>}\int_\ep^\infty dt\,e^{-tm^2}\,K_t(-\Th y.p,p)\,\delta^g_{z.p}(p').
$$
Now, one can qualitatively see in this distributional expression the
UV/IR entanglement phenomenon: thanks to the
estimate~(\ref{heatkernelestimate}), we have
\begin{align*}
\int_0^\infty dt\,e^{-tm^2}\,K_t(\Th y.p,p)&\leq
C\int_0^\infty dt\,\frac{e^{-tm^2}}{(4\pi t)^2}e^{-d_g^2(\Th y.p,p)/4(1+c)t}\\
&=\frac{C}{16\pi^2}\,\frac{4m\sqrt{1+c}}{d_g(\Th y.p,p)}\,
K_1\big(\frac{m\,d_g(\Th y.p,p)}{\sqrt{1+c}}\big)\\
&\sim C'\,d_g^{-2}(\Th y.p,p),
\quad y\to 0,
\end{align*}
and the reverse inequality also holds
$$
\int_0^\infty dt\,e^{-tm^2}\,K_t(\Th y.p,p)\geq  C''\,d_g^{-2}(\Th y.p,p)
$$
which points precisely to the UV/IR mixing, since $y\in\widehat{\R^l}$
has to be interpreted as a momentum.

For the two-point part of $\Ga_{1l,NP}^\ep[\vf]$ we have
\begin{align*}
\left|\Ga_{1l,NP,2P}^\ep[\vf]\right|\leq&
\frac{C\,\lambda}{12(2\pi)^l}\,\sup_{p\in M}\left\{\int_{\R^l} \,
d^lz\left|\vf(z.p)\right|\right\}\\
&\hspace{2cm}\times\int_\ep^\infty dt\,\frac{e^{-tm^2}}{(4\pi t)^2}\,
\int_M\mu_g(p)\,\left|\vf(p)\right|\int_{\R^l}\,d^ly\,
e^{-d_g^2(-\Th y.p,p)/4(1+c)t}\\
\leq&
\frac{C\,\lambda}{12(2\pi)^l}\,\sup_{p\in M}\left\{\int_{\R^l} \,
d^lz\left|\vf(z.p)\right|\right\}\|\vf\|_1\\
&\hspace{2cm}\times\sup_{p\in \supp(\vf)}
\left\{\int_\ep^\infty dt\,\frac{e^{-tm^2}}{(4\pi t)^2}\,
\int_{\R^l}\,d^ly\,
e^{-d_g^2(-\Th y.p,p)/4(1+c)t}\right\}.
\end{align*}
By the properness of $\a$, $\int_{\R^l} \, d^lz\left|\vf(z.p)\right|$
is finite for all $p\in M$ since $\{z\in\R^l: z.p\in\supp(\vf)\}$ is
compact for each $p\in M$ because $\vf$ has compact support. Thus,
$\tilde{\vf}(p):=\int_{\R^l} \, d^lz\left|\vf(z.p)\right|$ is constant and finite
on each orbit of $\a$, and if we denote $\pi:M\to M/\R^l$ the
projection on the orbit space, then $\tilde{\vf}$ factors through
$\pi$ to give a map $\bar{\vf}$ defined by
$\bar{\vf}(\pi(p)):=\tilde{\vf}(p)$. Finally,
$\bar{\vf}\in\Coo_c(M/\R^l)$ because if $p\notin\R^l.\supp(\vf)$, so
that $\pi(p)$ is not in the compact set $\pi(\supp(\vf))$, then
$\bar{\vf}(\pi(p))=0$. This proves that $\sup_{p\in
M}\left\{\int_{\R^l} \, d^lz\left|\vf(z.p)\right|\right\}<\infty$.
Furthermore, since $\a$ acts isometrically the induced metric
$\tilde{g}$ on the orbits (which are closed submanifolds since the action
is proper \cite{Michor}) is constant,
so
$$
d_g^2(y.p,p)=\sum_{i,j=1}^l\tilde{g}_{ij}(p)y^iy^j.
$$
Here, $\tilde{g}_{ij}(p)$ (which depend only on the the orbit of $p$)
are strictly positive continuous functions since in the non-periodic
case the action is free, and then $\{(0,p)\in \R^l\times M\}$ is the
only set for which $F(y,p):=d_g( y.p,p)$ vanish. Note that we can use
a global coordinate system (on one orbit) given by a suitable basis of
$\R^l$ in such a way that $\tilde{g}_{ij}(p)$ is diagonal. Thus, with
$\th:=(\det\Th)^{1/l}$, we have:
$$
\int_{\R^l}\,d^ly\,
e^{-d_g^2(-\Th y.p,p)/4(1+c)t}=
\left(\frac{4\pi(1+c) t}{\th^2}\right)^{l/2}(\det\tilde{g}(p))^{-1/2}.
$$
Hence, one obtains
$$
\left|\Ga_{1l,NP,2P}^\ep[\vf]\right|\leq
\frac{\lambda}{6}C(l,\tilde{g},\vf,\vf)\,\th^{-l}
\int_\ep^\infty dt\, t^{l/2-2}e^{-tm^2},
$$
where
\begin{align*}
C(l,\tilde{g},\vf_1,\vf_2):=&\\
&\hspace{-0.9cm}\frac{C(4\pi)^{l/2-2}(1+c)^{l/2}}{2(2\pi)^l}\|\vf_1\|_1
\sup_{p\in M}\left\{
\int_{\R^l}d^lz\left|\vf_2(z.p)\right|\right\}
\sup_{p\in\supp(\vf_1)}\left\{(\det\tilde{g}(p))^{-1/2}\right\}.
\end{align*}
Four the four-point part, similar estimates read:
$$
\left|\Ga_{1l,NP,4P}^\ep[\vf]\right|\leq
\frac{\lambda^2}{72}\,\th^{-l}\,\Big(3C(l,\tilde{g},\vf\Mop\vf,\vf\Mop\vf)
+4C(l,\tilde{g},\vf,\vf\Mop\vf\Mop\vf)\Big)
\int_\ep^\infty dt\, t^{l/2-1}e^{-tm^2}.
$$

We then have proved the following:
\begin{thm}
When $M$ is non-compact, satisfying all
assumptions on the behavior of the geometry at infinity
displayed above and endowed with a smooth proper
isometric action of $\R^l$, then
for $\vf\in\Coo_c(M)$ we have:
\item{i)}
$$
\left|\Ga_{1l,NP,2P}^\ep[\vf]\right|\leq\begin{cases}
   C_1(\vf,\Th)  &\text{for $l=4$}, \\
   C_2(\vf,\Th)|\ln\ep| &\text{for $l=2$}, \end{cases}
$$
\item{ii)}
$$
\left|\Ga_{1l,NP,4P}^\ep[\vf]\right|\leq
   C_3(\vf,\Th) \, \,\, \,\text{for $l=4$ or $l=2$}.
$$
\end{thm}
\noindent
The possible remaining divergence for $l=2$ refers to the fact that the
IR~singularity might be not integrable, as illustrated previously.
In this case, the two-point non-planar Green function does not
define a distribution and the theory is not renormalizable by
addition of local counter-terms, already in its one-loop approximation
order.

\section{Periodic deformations}

Periodic deformations (when the kernel of $\a$ is an integer lattice) behave
rather differently from non-periodic ones. In the following, we consider
$\ker\a=\beta \Z^l$ with $\beta$ a $l\times l$ integer matrix of rank $l$, so
that $\R^l/\beta\Z^l=:\T_\beta^l$ is compact. For the sake of simplicity, we
will often suppress the subscript $\beta$. \newline
Momentum space (the dual group of
$\T_\beta^l$) being discrete, IR~problems only occur for some values of the
momentum. In favorable cases one can extract the divergent field
configurations in the non-planar part (which are often finite in number when
$(2\pi)^{-1}\Th$ has irrational entries) and renormalize them like the planar
contributions; then there is no really UV/IR mixing. When
$(2\pi)^{-1}\Th$ has rational entries, the theory is equivalent to the undeformed
one, in the sense that there are infinitely many divergent field
configurations.

Although in all periodic cases we have a Peter--Weyl decomposition for
fields, only in the compact manifold case shall we be able to describe
the individual behavior of non-planar ``Feynman graphs'', defined
through that isotypic decomposition. Both in the compact and in the
non-compact case, by means of the off-diagonal heat kernel
estimate~(\ref{heatkernelestimate}), we show in the second subsection
how, for periodic deformations, the arithmetical nature of the entries
of $\Th$, more precisely, the existence or nonexistence of a
Diophantine condition on $\Th$, plays a role in determining the
analytical nature of $\Ga_{1l,NP}^\ep[\vf]$.

\subsection{Periodic compact case and the
individual behavior of non-planar graphs}

Because everything is explicit, we look first at the flat compact
case. Let $M=\T^4$ with the flat metric, let $\R^4$ act on it by
rotation (so and $l=4$ and we are in the `fully noncommutative
picture'). With the orthonormal basis
$\left\{\frac{e^{i<k,x>}}{(2\pi)^2}\right\}_{k\in\Z^4}$ of $L^2(\T^4,d^4x)$
the heat kernel is written
$$
K_t(x,y)=(2\pi)^{-4}\sum_{k\in\Z^4}e^{-t|k|^2}e^{i<k,x-y>},
$$
and we have
$$
e^{i<k,x>}\Mop e^{i<q,x>}=e^{-\sihalf \Th(k,q)}\,e^{i<k+q,x>},
$$
with $\Th(k,q):=\langle k,\Th q\rangle$. Expanding the background field $\vf$ in
Fourier modes $\vf=\sum_{k\in\Z^4}c_k\,e^{i<k,x>}$, with
$\{c_k\}_{k\in\Z^4}\in\SS(\Z^4)$ whenever $\vf\in\Coo(\T^4)$, we obtain:
\begin{align*}
\Ga_{NP}^\ep[\vf]=&\frac{1}{2}\sum_k
\frac{e^{-\ep(m^2+|k|^2)}}{m^2+|k|^2}\Big\lbrace
\frac{\lambda}{3!}\sum_r \,c_r\,c_{-r}\,e^{i\Th(k,r)}
   -\frac{\lambda^2}{2(3!)^2}
\frac{1}{m^2+|k|^2}\\
&\hspace{1cm}\times\sum_{r,s,u}\,c_r\,c_s\,c_{u-s}\,c_{-r-u}
   e^{-\sihalf\Th(r+s,u)}\Big(3\,e^{i\Th(k,r+s)}
+4\,e^{i\Th(k,r+u)}\Big)\Big\rbrace+O(\ep^0).
\end{align*}
We can now analyze the individual behavior of non-planar Feynman
diagrams.  One sees that, thanks to the phase factors, the sum over
$k$ is finite when $\ep$ goes to zero, whenever $(2\pi)^{-1}\Th$ has
irrational entries and $r\ne 0$ for the two-point part, or $r+s\ne 0$
and $r+u\ne 0$ for the four-point part.  In effect, returning to the
Schwinger parametrization (which exchanges large momentum divergences
with small-$t$ ones) and applying the Poisson summation formula with
respect to the sum over~$k$ we get:
$$
\sum_{k\in\Z^4}\frac{e^{i \Th(k,r)}}{m^2+|k|^2}=
\sum_{k\in\Z^4} \int_0^\infty dt\,
\frac{e^{-tm^2}}{(4\pi t)^{2}}\,e^{-|2\pi k-\Th r|^2/4t}.
$$
Hence, the $t$-integral is finite whenever $r\ne 0$ and $\frac{\Th
r}{2\pi}\notin \Q^l$. Essentially the same conclusion holds for the
four-point part.

We now go to the general periodic compact case. In order to be able to
calculate, we make explicit use of the invariance of the heat kernel
under $\a$. Let us decompose $\H=L^2(M,\mu_g)$ in spectral
subspaces with respect to the group action:
$$
\H=\bigoplus_{k\in\Z^l}\H_k.
$$
Each $\H_k$ is stable under $V_z$ (recall that $V_z$ denotes the
induced action on $\H$) for all $z\in\R^l$; and furthermore all
$\psi\in\H_k$ satisfy $V_z\psi=e^{-i<z,k>}\psi$. Note that if
$\psi\in\H_k$ then $|\psi|\in\H_0$. Let $P_k$ be the orthogonal
projection on $\H_k$. Because the Laplacian commutes with $V_z$, the
heat operator also commutes with $P_k$; hence $e^{-t\tri}$ is block
diagonalizable with respect to the decomposition
$\H=\bigoplus_{k\in\Z^l}\H_k$:
$$
e^{-t\tri}=\sum_{k\in\Z^l}\,P_k\,e^{-t\tri}\,P_k.
$$
In each $\H_k$ the operator $0\leq P_k\,e^{-t\tri}\,P_k$ is
trace-class, so it can be written as
$$
P_k\,e^{-t\tri}\,P_k=\sum_{n\in\N}
e^{-t\lambda_{k,n}}|\psi_{k,n}\rangle\langle\psi_{k,n}|,
$$
where $\{\psi_{k,n}\}_{n\in\N}$ is an orthonormal basis of $\H_k$
consisting of eigenvectosr of $P_k\tri P_k$ with eigenvalue
$\lambda_{k,n}$. The heat semigroup being Hilbert-Schmidt, its kernel
can be written as a ($L^2(M\times M,\mu_g\times\mu_g)$-convergent)
sum:
\begin{equation}
\label{eq:hk}
K_t(p,p')=\sum_{k\in\Z^l}\sum_{n\in\N}
e^{-t\lambda_{k,n}}\psi_{k,n}(p)\overline{\psi_{k,n}}(p').
\end{equation}
Because each $\psi_{k,n}(p)$ lies in $\H_k$, the invariance
property (\ref{invprop}) $K_t(z.p,z.p')=K_t(p,p')$ is explicit.

Any $\vf\in C^\infty(M)$ has a Fourier decomposition $\vf=
\sum_{r\in\Z^l}\vf_r$, such that $\{\|\vf_r\|_\infty\}\in\SS(\Z^l)$
and $\a_z(\vf_r) =e^{-i<z,r>}\vf_r$.  Furthermore, this decomposition
provides a notion of Feynman diagrams, that is of amplitude associated
with a fixed field configuration.  The non-planar one-loop regularized
effective action reads:
\begin{align*}
\Ga_{1l,NP}^\ep[\vf]=&\frac{1}{2}\int_M\mu_g(p)\sum_{k\in\Z^l}\sum_{n\in\N}
\frac{e^{-\ep(m^2+\lambda_{k,n})}}
{m^2+\lambda_{k,n}}\, |\psi_{k,n}|^2(p)\Big\lbrace
\frac{\lambda}{3!}\sum_{r,s\in\Z^l} \,
\vf_r(p)\,\vf_s(p)\,e^{-i\Th(k,s)} \\
&-\frac{\lambda^2}{2(3!)^2}
\frac{1}{m^2+\lambda_{k,n}}\sum_{r,s,u,v\in\Z^l}\,
\vf_r(p)\,\vf_s(p)\,\vf_u(p)\,\vf_v(p)\\
&\hspace{1cm}\times\Big(3\,e^{-\sihalf(\Th(r,s)+\Th(u,v))}
e^{-i\Th(k,u+v)}
+4\,e^{-\sihalf \Th(r+s,u+s)}
e^{-i\Th(k,v)}\Big)\Big\rbrace+O(\ep^0).
\end{align*}
Although we do not know the explicit form of the $\psi_{k,n}$, we can
by momentum conservation reduce the sums exactly as in the NC torus
case, as shown in the following lemma.
\begin{lem}[Momentum conservation]
\label{lem:u}
Let $\psi_i\in\H_{k_i}\cap L^q(M,\mu_g)$ for $i=1,\dots,q$. Then:
$$
\int_M\mu_g\,\psi_1\cdots\psi_q=
C(\psi_1,\cdots,\psi_q)\,\delta_{k_1+\cdots +k_q,0}.
$$
\end{lem}
\begin{proof}
By the $\a$-invariance of $\mu_g$ and with the relation
$\a_z(\psi_i)=e^{-i<z,k_i>}\psi_i$ we have
$$
\int_M\mu_g\,\psi_1\cdots\psi_q=
e^{i<z,k_1+\cdots+k_q>}\int_M\mu_g\,\psi_1\cdots\psi_q,
$$
for all $z\in\R^l$; the result follows.
\end{proof}

Because $|\psi_{k,n}|^2(p)$ is constant
on the orbits of $\a$ and $\vf_r\in\Coo(M)\subset L^q(M,\mu_g)$
for all $q\geq1$, Lemma \ref{lem:u} gives
\begin{align}
\label{trs}
\Ga_{1l,NP}^\ep[\vf]=&\frac{1}{2}\int_M\mu_g(p)\sum_{k\in\Z^l}\sum_{n\in\N}
\frac{e^{-\ep(m^2+\lambda_{k,n})}}{m^2+\lambda_{k,n}}\,
|\psi_{k,n}|^2(p)\Big\lbrace
\frac{\lambda}{3!}\sum_{r\in\Z^l} \,\vf_r(p)\,
\vf_{-r}(p)\,e^{i<k,\Th r>} \nonumber\\
&-\frac{\lambda^2}{2(3!)^2}
\frac{1}{m^2+\lambda_{k,n}}\sum_{r,s,u\in\Z^l}\,
\vf_r(p)\,\vf_s(p)\,\vf_{u-s}(p)\,\vf_{-r-u}(p)\nonumber\\
&\hspace{2cm}\times e^{-\sihalf\Th(r+s,u)}\Big(3\,e^{i\Th(k,r+s)}
+4\,e^{i\Th(k,r+u)}\Big)\Big\rbrace+O(\ep^0).
\end{align}
To analyze the divergences when $\ep\rightarrow 0$ for a fixed field
configuration, note that if we re-index $\lambda_{k,n}$ in a standard
way ($\lambda_0\leq\cdots\leq\lambda_n\leq\cdots$ ), Weyl's estimate
asserts that $\lambda_n\sim n^{1/2}$, hence
$$
\sum_{k\in\Z^l}\sum_{n\in\N}\frac{|\psi_{k,n}|^2(p)}{(m^2+\lambda_{k,n})^N}
=\sum_{n\in\N}\frac{|\psi_n(p)|^2}{(m^2+\lambda_n)^N}
=K_{(m^2+\tri)^{-N}}(p,p),
$$
is finite if and only if $N>2$. We see that the sum over $n$ and $k$
in~(\ref{trs}) diverges in the limit $\ep\rightarrow 0$ for certain
values of the momenta ($r=0$ for the two-point part, $r+s=0$ and
$r+u=0$ for the four-point part) if $(2\pi)^{-1}\Th$ has irrational entries.
When the entries $(2\pi)^{-1}\Th$ are rational, there are infinitely many
divergent field configurations since $e^{-i<k,\Th r>}=1$  for infinitely
many $k$ whenever $\frac{\Th r}{2\pi}\in\Q^l$. For other configurations,
convergence is guaranteed by the estimate~(\ref{heatkernelestimate}), as
shown in the next subsection.

In summary, we have shown that the behavior of an individual field
configuration in the non-planar sector for any periodic compact
deformation reproduces the main features of the noncommutative torus.

In the next paragraph, the arithmetic nature of the entries of $\Th$
gets into the act; also we show there that the possible existence of
fixed points for the action may give rise to additional divergences.

\subsection{General periodic case and the Diophantine condition}

Assume now that $\a$ periodic, but $M$ can be compact or not (within
the hypothesis of section 4.2 when $M$ is not compact).  In this
general setup, the Peter-Weyl decomposition still exists, but the heat
operator, not being a priori compact, cannot be written
as~(\ref{eq:hk}).  Thus we return to the off-diagonal heat kernel
estimate.  In this case, using Lemma \ref{lem:u} and the
$\a$-invariance of $K_t$, we obtain:
\begin{align*}
\Ga_{1l,NP}^\ep[\vf]=&\frac{1}{2}
\int_\ep^\infty\,dt\,e^{-tm^2}\int_M\mu_g(p)\Big\lbrace
\frac{\lambda}{3!}\sum_{r\in\Z^l}K_t(\Th r.p,p)\vf_r(p) \,\vf_{-r}(p)\\
&-\frac{t\,\lambda^2}{2(3!)^2}
\sum_{r,s,u\in\Z^l}\,\vf_r(p)\,\vf_s(p)\,\vf_{u-s}(p)\,\vf_{-r-u}(p)
\,\,e^{-\sihalf \Th(r+s,u)}\\
&\hspace{2cm}\times \Big(
3\,K_t(\Th(r+s).p,p)
+4\,K_t(\Th (r+u).p,p)\Big)\Big\rbrace
+O(\ep^0).
\end{align*}
We consider only the case $(2\pi)^{-1}\Th$ has irrational entries, from now
on.  Then divergences appear when $r=0$ for the two-point function and
$r+s=0,r+u=0$ for the four-point functions. This leads us to introduce a
\emph{reduced} non-planar one-loop effective action
$\Ga_{1l,NP}^{\ep,red}[\vf]$ by subtracting the divergent field
configurations; for renormalization purposes, they have to be treated
together with the planar sector.
\begin{align*}
\Ga_{1l,NP}^{\ep,red}[\vf]:=&\frac{1}{2}
\int_\ep^\infty\,dt\,e^{-tm^2}\int_M\mu_g(p)\Big\lbrace
\frac{\lambda}{3!}\tsum' K_t(\Th r.p,p)\vf_r(p) \,\vf_{-r}(p)\\
&\hspace{2cm}-\frac{t\,\lambda^2}{2(3!)^2}
\tsum'\,\vf_r(p)\,\vf_s(p)\,\vf_{u-s}(p)\,\vf_{-r-u}(p)
e^{-\sihalf \Th(r+s,u)}\\
&\hspace{4cm}\times\Big(
3\,K_t(\Th(r+s).p,p)+4\,K_t(\Th (r+u).p,p)
\Big)\Big\rbrace.
\end{align*}
Here $\tsum'$ is the notation for $\sum_{r\in\Z^l,\,r\ne 0}$ in the
two-point part, $\sum_{r,s,u\in\Z^l,\,r+s\ne 0}$ and\newline
$\sum_{r,s,u\in\Z^l,\,r+u\ne 0}$ in respectively the first and second
piece of the four-point part.\ Using now the
estimate~(\ref{heatkernelestimate}) and performing the
$t$-integration, we obtain:
\begin{align}
\label{belll}
\lim_{\ep\to 0}\left|\Ga_{1l,NP}^{\ep,red}[\vf]\right|
\leq& \frac{C}{32\pi^2}\int_M\mu_g(p)\Big\lbrace
\frac{\lambda}{3!}\tsum'|\vf_r(p) |\,|\vf_{-r}(p)|
\frac{4m\sqrt{1+c}}{d_g(\Th r.p,p)} K_1\big(\frac{m\,d_g(\Th
r.p,p)}{\sqrt{1+c}}\big)\nonumber\\
&+\frac{\lambda^2}{2(3!)^2}\tsum'\,
|\vf_r(p)|\,|\vf_s(p)|\,|\vf_{u-s}(p)|\,|\vf_{-r-u}(p)|\nonumber\\
&\times\Big(3K_0\big(\frac{m\,d_g(\Th(r+s).p,p)}{\sqrt{1+c}}\big)+
4K_0\big(\frac{m\,d_g(\Th(r+u).p,p)}{\sqrt{1+c}}\big)\Big)\Big\rbrace.
\end{align}

\begin{defn}
$\th\in\R^l\setminus\mathbb{Q}^l$
satisfies a Diophantine condition if there exists $C>0$, $\beta\geq0$
such that for all $n\in\Z^l_{\setminus\{0\}}$:
$$
\|n\th\|_{\T^l}:=\inf_{k\in\Z^l}|n\th+k|\geq C|n|^{-(l+\beta)}.
$$
\end{defn}
Diophantine conditions constitute a way to characterize and classify
irrational numbers which are ``far from the rationals'' in the sense
of being badly approximated by rationals.  The set of numbers
satisfying a Diophantine condition is `big' (of full Lebesgue measure)
in the sense of measure theory, but `small' (of first category) in the
sense of category theory~\cite{Oxtoby}.  Again
because the metric is constant on the orbits we have:
$$
d_g^2(y.p,p)=\inf_{k \in\Z^l}\left(\sum_{i,j=1}^l
\tilde{g}_{ij}(p)(y^i+k^i)(y^j+k^j)\right).
$$
Recall also that the modified Bessel functions have the following
behavior near the origin
$$
K_1(x)=\frac{1}{x}+O(x^0),\hspace{1cm}K_0(x)=
-\gamma+\ln(2)-\ln(x)+O(x),
$$
where $\gamma$ is the Euler constant.  Thus, in view of
$\{\|\vf_r\|_\infty\}\in\SS(\Z^l)$, and provided the integral over the
manifold with the measure $\mu_g$ can be carried out, in~(\ref{belll}) we
have convergence if and only if $d_g^{-2}(\Th r.p,p)\in\SS'(\Z^l)$,
that is, if and only if the entries of $\Th$ satisfy a Diophantine
condition.  This result seems to be new, although the pertinence of
Diophantine conditions in NCQFT had been conjectured by Connes long
ago.  Recently, these conditions have been found to play a role in
Melvin models with irrational twist parameter in conformal field
theory~\cite{KMM}.

\smallskip

We said above: ``provided the integral over the manifold with the measure
$\mu_g$ can be carried out''.  This because $d_g^{-2}(\a_y(.),.)$ for a
non-zero $y\in\T^l$ might not be locally integrable with respect to the
measure given by the Riemannian volume form.  Problems may appear on a
neighborhood of the set of points with non-trivial isotropy groups.  In fact,
by simple dimensional analysis, we expect serious trouble when the isotropy
group is one dimensional.  For $p\in M$ let $H_p$ its isotropy group and let
$M_{sing}:=\{p\in M:H_p\ne\{0\}\}$.  Recall that $M_{sing}$ is closed and of
zero-measure in $M$ since the action is proper (see \cite{Michor}), and note
that for a non-zero $y\in\T^l$, $d_g(y.p,p)=0$ if and only if $p\in M_{sing}$
and $y\in H_p$.  On $M_{reg}:=M\setminus M_{sing}$ (the set of principal
orbit type), since the action is free, one can define normal coordinates on a
tubular neighborhood of an orbit $\T^l.p$.  Let $(\hat{x}^\mu,\tilde{x}^i)$,
$\mu=1,\cdots,n-l$, $i=1,\cdots,l$ be respectively the transverse and the
torus coordinates of a point $p\in M_{reg}$.  Because the action is isometric,
in this coordinate system the metric takes the form
$$
g(\hat{x},\tilde{x})= \begin{pmatrix} h(\hat{x}) & l(\hat{x})\\
l(\hat{x}) & \tilde{g}(\hat{x}) \end{pmatrix},
$$
where $\tilde{g}$ is the induced (constant) metric on the orbit.  Such
coordinate system is singular with singularities located at each point
of $M_{sing}$, and when $x\equiv (\hat{x}^\mu,\tilde{x}^i)$ approach
$p_0\in M_{sing}$, $\tilde{g}(\hat{x})$ collapses to a
$l-dim(H_{p_0})$ rank matrix.  Since in this coordinate system
$\mu_g(p)\,d^{-2}_g(y.p,p)$ equals
$$
\frac{\sqrt{\det
g(\hat{x})}}{\sum_{i,j=1}^l\tilde{g}_{ij}(\hat{x})y^iy^j}
d^l\tilde{x}\,d^{n-l}\hat{x},
$$
when $dim(H_{p_0})=1$ the singularity of $d^{-2}_g(y.p,p)$ for $p\to p_0$
cannot be cancelled by $\sqrt{\det g}$.  This is a new feature of the UV/IR
mixing for generic periodic isospectral deformations which needs to be
investigated in detail in each model; it occurs, for instance, for the
Connes--Landi spheres and their ambient spaces.  Let us summarize:
\begin{thm}
For $M$ compact or not (within the assumptions displayed in section 4.2 in the
non-compact case), endowed with a smooth isometric action of the compact group
$\T^l$, $l=2$ or $l=4$ and with a deformation matrix whose entries satisfy a
Diophantine condition, then for any external field $\vf\in\Coo_c(M,)$ vanishing
in a neighborhood $M_{sing}$ the one-loop non-planar reduced effective action
is finite.
\end{thm}

In other words, if the Diophantine condition is not satisfied or if
$d^{-2}_g(\a_y(.),.)\notin L^1_{loc}(M,\mu_g)$ then the reduced
non-planar two-point function does not define a distribution and
the theory is not renormalizable, already at one-loop, by addition
of local counter-terms.

\section{Summary and perspectives}

We have shown the existence of the UV/IR mixing for isospectral
deformations of curved spaces.

For periodic deformations the entanglement only concerns (at the level
of the two-point function) the 0-th component of the field in the
spectral subspace decomposition induced by the torus action.  In this
case, the UV/IR mixing does not generate much trouble since one can
treat it for renormalization purposes together with the planar sector.

In the non-periodic situation, we obtain non-planar Green functions
which present the mixing in a similar form to the Moyal plane
paradigms.

Our approach gives an algebraic way to understand the presence of the
non-planar sector for those theories: it comes from the product of
left and right regular representation operators.  As a byproduct of
our trace computations, we obtain that the better behavior of the
non-planar sector is due to the presence of the off-diagonal heat
kernel in the integrals.

However, its regularizing character depends highly on the geometric data.  For
non-periodic deformations, the conclusion is that when the noncommutative
rank is equal to two, the non-planar 1PI two-point  Green function does not
define a distribution and the associated effective action remains
divergent~\cite{Parthenope}.  Only the group action of rank~four gives rise
to a UV divergent-free non-planar sector in the 4-dimensional manifold case.
When the action is periodic, we have shown that it is necessary that the
entries of $(2\pi)^{-1}\Th$ satisfy a Diophantine condition to ensure
finiteness of the reduced non-planar effective action, i.e. in order that
the reduced non-planar 1PI two-point Green function define a
distribution.  Additional divergences may exist due to the possible
fixed points structure of the action $\a$.

Our treatment of the generic UV/IR behavior, can be generalized to higher
dimensional isospectral deformations and/or to gauge theories. Also, we have
restricted ourselves to the 4-dimensional case, for the sake of simplicity and
physical interest, but it is clear that the heat kernel techniques employed
here apply to higher dimensional scalar theories.

\smallskip

For gauge theory on (any dimensional) isospectral deformations
manifolds, there is an intrinsic way to define noncommutative actions
of the Yang--Mills type.  For any $\omega\in\Omega^p(M),\eta\in\Omega^q(M)$
(say compactly supported and smooth with respect to $\a$) one can set
$$
\omega\wedge_\Th\eta:=(2\pi)^{-l}\int_{\R^{2l}}\,d^ly\,d^lz\;e^{-i<y,z>}\;
(\a^*_{-\thalf\Th y}\omega)\wedge(\a^*_{z}\eta),
$$
where $\a^*_z$ is the pull-back of $\a_z$ on forms.  Given now an
associated vector bundle $\pi:E\rightarrow M$ with compact structure
group $G\subset U(N)$, and a connection $A\in\Omega^1(M,\Lie(G))$ we
define the NC analogue of the YM action
$$
S_{YM}(A):=\int_M\tr(F_\Th\wedge_\Th \ast_{_H} F_\Th),
$$
where $F_\Th:=dA+A\wedge_\Th A$.  In this
context, one can prove a trace property, namely:
$$
\int_M\omega\wedge_\Th\ast_{_H}\eta=
\int_M\omega\wedge\ast_{_H}\eta,\,\forall \omega,\eta\in \Omega^p(M).
$$
Hence $S_{YM}(A)$ equals $\int_M\tr(F_\Th\wedge \ast_{_H} F_\Th)$.  To
manage the quantization, one can once again use the background field
method in the background gauge, and if we ignore the Gribov ambiguity,
the one-loop effective action reduce to the computation of
determinants of operators (quadratic part in $A$ of $S_{YM}+S_{gf}$
and Faddeev--Popov determinant) which can be locally expressed as
$$
(\nabla_\mu+L_{A_\mu}-R_{A_\mu})(\nabla^\mu+L_{A^\mu}-R_{A^\mu})+B,
$$
where $B$ is bounded and contains left, right and a product of left
and right twisted multiplication operators.  It is then clear that
UV/IR mixing will appears in the same form as in the flat situations
(see \cite{KW, MR, MRS}).

A further interesting task is be to look at what happens for a
Grosse--Wulkenhaar like model for the non-compact case.  In \cite{GW}
it is proved that if we add a confining potential (harmonic oscillator
in their work) in the usual $\lambda \vf^{\Mop 4}$ theory on the four
dimensional Moyal plane, i.e. the Grosse--Wulkenhaar action
\begin{align*} S_{GW}[\vf]:=&\int
d^4x\,\Bigl[\thalf(\pa_\mu\vf\mop\pa^\mu\vf)(x)
+2\frac{\Omega^2}{\th^2}(x_\mu\vf)\mop(x^\mu\vf)\nonumber\\
&\hspace{4,4cm}+\frac{m^2}{2}\vf\mop\vf(x)
+\frac{\lambda}{4!}\vf\mop\vf\mop\vf\mop\vf(x)\Bigr],
\end{align*}
then the theory is perturbatively renormalisable to all orders
in~$\lambda$.  The deep meaning of this result is not yet fully
understood, but some explanations can be mentioned.  First, to add a
confining potential is in some sense equivalent to a compactification
of the Moyal plane and in the second hand, the particular choice of
the potential corresponds to a Moyal-deformation of both the
configuration and the momentum space.  This can be seen by the
invariance (up to a rescaling) of this action under
$p_\mu\leftrightarrow 2(\th^{-1})_{\mu\nu}x^\nu$,
$\widehat{\vf}(p)\leftrightarrow(\pi\th)^2\vf(x)$.  This point needs
to be clarified.  It would be good to know whether their
renormalizability conclusion (UV/IR decoupling) holds in the general
context when one adds a coupling with a confining potential in the
scalar theory.

Last, but not least, it remains to see whether the UV/IR entanglement
concerns only $\th$-deformations or not.  Connes--Dubois-Violette
\cite{ConnesDV} 3-spheres and 4-planes, whose defining algebras are
related to Sklyanin algebras, are good candidates to test this point.

\section*{Acknowledgements}
I am very grateful to J. M. Gracia-Bond\'{\i}a, J.~C. V\'arilly and my advisor
B.~Iochum for their help.  I also would like to thank M.~Grasseau,
T.~Krajewski, F.~Ruiz~Ruiz and R.~Zentner for fruitful discussions and/or
suggestions.  Special thanks are also due to the Departamento de
F\'{\i}sica Te\'orica I of the Universidad Complutense de Madrid for its
hospitality during the final stages of this work. I finally would like to thank
the Referee for his enlightened remarks.


\begin{thebibliography}{35}

\bibitem{Braunss}
G. Braunss,
``On the regular Hilbert space representation of a Moyal quantization'',
J. Math. Phys. {\bf 35} (1994), 2045--2056.

\bibitem{Chavel}
I. Chavel,
\emph{Eigenvalues in Riemannian Geometry},
Academic Press, London and San Diego, 1984.

\bibitem{CR}
I. Chepelev and R. Roiban,
``Renormalization of Quantum Fields Theories on
Noncommutative $\R^d$. I. Scalar'', J. High Energy Phys. {\bf 5} (2000), 137--168.

\bibitem{Book}
A. Connes,
\emph{Noncommutative Geometry},
Academic Press, London and San Diego, 1994.

\bibitem{Connesgrav}
A. Connes,
``Gravity coupled with matter and the foundation of noncommutative geometry'',
Commun. Math. Phys. {\bf 182} (1996), 155--176.

\bibitem{ConnesLa}
A. Connes and G. Landi,
``Noncommutative manifolds, the instanton algebra and isospectral
deformations'',
Commun. Math. Phys. {\bf 221} (2001), 141--159.

\bibitem{ConnesDV}
A. Connes and M. Dubois-Violette,
``Noncommutative finite-dimensional manifolds. I. Spherical manifolds
and related examples'',
Commun. Math. Phys. {\bf 230} (2002), 539--579.

\bibitem{CRTN}
T. Coulhon, E. Russ and V. Tardivel--Nachef,
``Sobolev algebras on Lie groups and Riemannian manifolds'',
Amer. J. Math. {\bf 123} (2001), 283--342.

\bibitem{Davies}
E. B. Davies,
\emph{Heat Kernels and Spectral Theory},
Cambridge University Press, Cambridge, 1989.

\bibitem{DN}
M. R. Douglas and N. A. Nekrasov,
``Noncommutative Fields Theory''
Rev. Modern Phys. {\bf 73} (2001), 977--1024.

\bibitem{Filk}
T. Filk, ``Divergences in a Field Theory on Quantum Space'',
Phys. Lett. B {\bf 376} (1996), 53--58.

\bibitem{Himalia}
V. Gayral, J. M. Gracia-Bond\'{\i}a, B. Iochum, T. Sch\"ucker and
J.~C. V\'arilly,
``Moyal planes are spectral triples'',
Commun. Math. Phys. {\bf 246} (2004), 569--623.

\bibitem{GI}
V. Gayral and B. Iochum,
``The spectral action for Moyal planes'',
hep-th/0402147. To appear in J. Math. Phys.

\bibitem{Parthenope}
V. Gayral, J. M. Gracia-Bond\'{\i}a and F. Ruiz Ruiz,
``Trouble with space and nonconstant noncommutativity field
theory", hep-th/0412235.

\bibitem{Melpomene}
V. Gayral, J. M. Gracia-Bond\'{\i}a and F. Ruiz Ruiz,
``Position-dependent noncommutative products: classical construction and field theory'',
hep-th/0504022

\bibitem{GIV}
V. Gayral, B. Iochum and J.~C. V\'arilly,
``Dixmier trace on non-compact isospectral deformations'',
in preparation.

\bibitem{Gilkey}
P. B. Gilkey, \emph{Invariance Theory, the Heat Equation and
the Atiyah--Singer Index Theorem},
2nd edition, CRC Press, Boca Raton, FL, 1995.

\bibitem{Carme}
J. M.
Gracia-Bond\'{\i}a,
``Improved Epstein--Glaser renormalization in
coordinate space I.
Euclidean framework'',
Math. Phys. Analysis Geom.
{\bf6} (2003), 59--88.

\bibitem{Polaris}
J. M. Gracia-Bond\'{\i}a, J. C. V\'arilly and H.
Figueroa,
\emph{Elements of Noncommutative Geometry},
Birkh\"auser
Advanced Texts, Birkh\"auser, Boston, 2001.

\bibitem{GW}
H. Grosse and R.
Wulkenhaar,
``Renormalisation of $\phi^4$-Theory on Noncommutative
$\R^4$ to all order'',
hep-th 0403232.

\bibitem{KW}
T. Krajewski and
R. Wulkenhaar,
``Perturbative Quamtum Gauge Fields on the
Noncommutative Torus'',
Int. J. Mod. Phys. A {\bf 15} (2000),
1011--1030.

\bibitem{KMM}
D. Kurasov, J. Marklof and G. W.
Moore,
``Melvin models and Diophantine approximation'',
hep-th
0407150.

\bibitem{MR}
C.P. Martin and D. Sanchez--Ruiz,
``The One
Loop UV Divergent stucture of $U(1)$ Yang--Mills Theory
on
Noncommutative $\R^4$'', Phys. Rev. Lett. {\bf 83} (1999),
476--479.

\bibitem{MRS}
S. Minwalla, M. V. Raamsdonk and N.
Seiberg,
``Noncommutative Perturbative Dynamics'',
J. High Energy
Phys., {\bf 2} (2000), 20--31.

\bibitem{Michor}
P. W.
Michor,
``Isometric actions of Lie groups and invariants'',
Notes of
a lecture course at the University of Vienna, July
1997.

\bibitem{MS}
S. B. Myers and N. Steenrod,
``On the group of isometries of a Riemannian manifold'',
Ann. Math. {\bf 40} (1939), 406--416.

\bibitem{Oxtoby}
J. C. Oxtoby,
\emph{Measure and
Category},
Springer, Berlin, 1972.

\bibitem{RieffelDefQ}
M. A.
Rieffel,
\emph{Deformation Quantization for Actions of
$\R^d$},
Memoirs Amer. Math. Soc. {\bf 506}, Providence, RI,
1993.

\bibitem{FRR}
F. Ruiz Ruiz,
``UV/IR mixing and the Goldstone
theorem in noncommutative field
theory'',
Nucl. Phys. {\bf B637}
(2002), 143--167.

\bibitem{NR}
O. Schnetz,
``Natural renormalization'',
J. Math. Phys. {\bf 38}
(1997) 738--758.

\bibitem{Sitarz}
A. Sitarz,
``Rieffel's deformation
quantization and isospectral deformations'',
Int. J. Theor. Phys.
{\bf 40} (2001), 1693--1696.
IJTPB,40,1693;

\bibitem{Z}
R. J. Szabo,
``Quantum Fields Theory on
Noncommutative space'',
Phys. Rep. {\bf 37} (2003),
207--299.

\bibitem{Larissa}
J. C. V\'arilly,
``Quantum symmetry
groups of noncommutative spheres'',
Commun. Math. Phys. {\bf 221}
(2001), 511--523.

\bibitem{Atlas}
J.
C. V\'arilly and J. M. Gracia-Bond\'{\i}a,
``On the ultraviolet behavior of
quantum fields over noncommutative manifolds'',
Int. J. Mod. Phys. {\bf A14} (1999), 1305--1323.

\bibitem{Vass}
D. V.
Vassilevich,
``Non-commutative heat kernel'',
Lett. Math. Phys. {\bf
67} (2004), 185--194.

\bibitem{ZJ}
J. Zinn-Justin,
\emph{Quantum
Field Theory and Critical Phenomena}, fourth edition,
Clarendon
Press, Oxford, 2002.










\end{thebibliography}
\end{document}